\documentclass[sigconf]{acmart}

\setcopyright{rightsretained}

\usepackage{etex}
\usepackage{subfigure}
\usepackage{graphicx}
\usepackage{amsmath}
\usepackage{bm}
\usepackage{url}
\usepackage{stmaryrd}
\usepackage{balance}
\usepackage{amssymb}
\usepackage{pifont}
\usepackage[usenames,dvipsnames]{pstricks}
\usepackage{epsfig}
\usepackage{epstopdf}
\usepackage{multirow}
\usepackage{booktabs}
\usepackage{pgffor}
\usepackage{tikz}
\usepackage{enumitem}
\usepackage{algorithm,algorithmic}
\usepackage{natbib}

\newcommand{\ch}{\textsc{Alph.}}
\newcommand{\nonch}{\textsc{Spec.}}
\newcommand{\accel}{\textsc{Accel.}}
\newcommand{\dmvm}{\textsc{DMVM}}
\newcommand{\dfm}{\textsc{DFM}}
\newcommand{\dnn}{\textsc{DNN}}
\newcommand{\xgb}{\textsc{XGB}}
\newcommand{\svm}{\textsc{SVM}}
\newcommand{\lr}{\textsc{LR}}

\begin{document}
\title{DeepMood: Modeling Mobile Phone Typing Dynamics for Mood Detection}
\author{Bokai Cao$^1$, Lei Zheng$^1$, Chenwei Zhang$^1$, Philip S. Yu$^{1,2}$, Andrea Piscitello$^1$, John Zulueta$^3$, \\Olu Ajilore$^3$, Kelly Ryan$^4$, and Alex D. Leow$^{1,3,5}$}
\affiliation{$^1$Department of Computer Science, University of Illinois at Chicago}
\affiliation{$^2$Institute for Data Science, Tsinghua University}
\affiliation{$^3$Department of Psychiatry, University of Illinois at Chicago}
\affiliation{$^4$Department of Psychiatry, University of Michigan}
\affiliation{$^5$Department of Bioengineering, University of Illinois at Chicago}
\email{caobokai, lzheng21, czhang99, psyu, apisci2@uic.edu}
\email{jzulueta, oajilore, aleow@psych.uic.edu}
\email{karyan@med.umich.edu}

\renewcommand{\shortauthors}{B. Cao et al.}

\begin{abstract}
The increasing use of electronic forms of communication presents new opportunities in the study of mental health, including the ability to investigate the manifestations of psychiatric diseases unobtrusively and in the setting of patients' daily lives. A pilot study to explore the possible connections between bipolar affective disorder and mobile phone usage was conducted. In this study, participants were provided a mobile phone to use as their primary phone. This phone was loaded with a custom keyboard that collected metadata consisting of keypress entry time and accelerometer movement. Individual character data with the exceptions of the backspace key and space bar were not collected due to privacy concerns. We propose an end-to-end deep architecture based on late fusion, named DeepMood, to model the multi-view metadata for the prediction of mood scores. Experimental results show that 90.31\% prediction accuracy on the depression score can be achieved based on session-level mobile phone typing dynamics which is typically less than one minute. It demonstrates the feasibility of using mobile phone metadata to infer mood disturbance and severity.
\end{abstract}




\fancyhead{}
\settopmatter{printccs=true, printacmref=false}

\copyrightyear{2017} 
\acmYear{2017} 
\setcopyright{acmcopyright}
\acmConference{KDD'17}{}{August 13--17, 2017, Halifax, NS, Canada}
\acmPrice{15.00}\acmDOI{10.1145/3097983.3098086}
\acmISBN{978-1-4503-4887-4/17/08}

\maketitle

\section{Introduction}

Mobile phones, in particular, ``smartphones'' have become near ubiquitous with 2 billion smartphone users worldwide. This presents new opportunities in the study and treatment of psychiatric illness including the ability to study the manifestations of psychiatric illness in the setting of patients' daily lives in an unobtrusive manner and at a level of detail that was not previously possible. Continuous real-time monitoring in naturalistic settings and collection of automatically generated smartphone data that reflect illness activity could facilitate early intervention and have a potential use as objective outcome measures in efficacy trials \cite{ankers2009objective,bopp2010longitudinal,faurholt2016behavioral}.

While mobile phones are used for a variety of tasks the most widely and frequently used feature is text messaging. To the best of our knowledge, no previous studies \cite{american2013diagnostic,puiatti2011smartphone,frost2013supporting,gruenerbl2014using,schleusing2011monitoring,valenza2014wearable} have investigated the relationship between mobile phone typing dynamics and mood states. In this work, we aim to determine the feasibility of inferring mood disturbance and severity from such data. In particular we seek to investigate the relationship between the digital footprints and mood in bipolar affective disorder which has been deemed the most expensive behavioral health care diagnosis \cite{peele2003insurance}, costing more than twice as much as depression per affected individual \cite{laxman2008impact}. For every dollar allocated to outpatient care for people with bipolar disorder, \$1.80 is spent on inpatient care, suggesting early intervention and improved prevention management could decrease the financial impact of this illness \cite{peele2003insurance}.

\begin{figure}[t]
\centering
\begin{minipage}[l]{\columnwidth}
\centering
\includegraphics[width=1\textwidth]{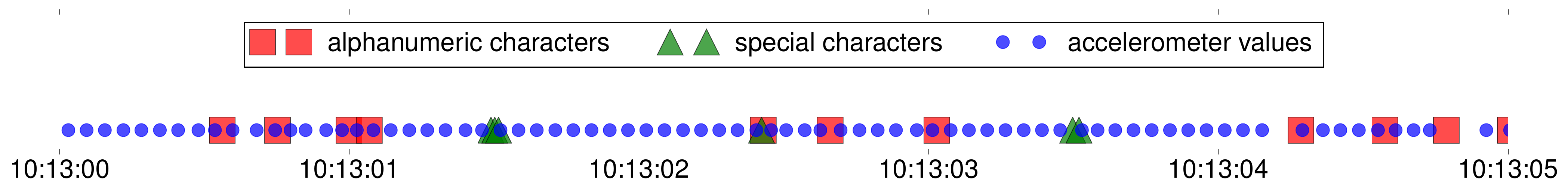}
\end{minipage}
\caption{A sample of the collected data in time series.}
\label{fig:timeline}
\end{figure}

\begin{figure*}[t]
\centering
\begin{minipage}[l]{1.6\columnwidth}
\centering
\includegraphics[width=1\textwidth]{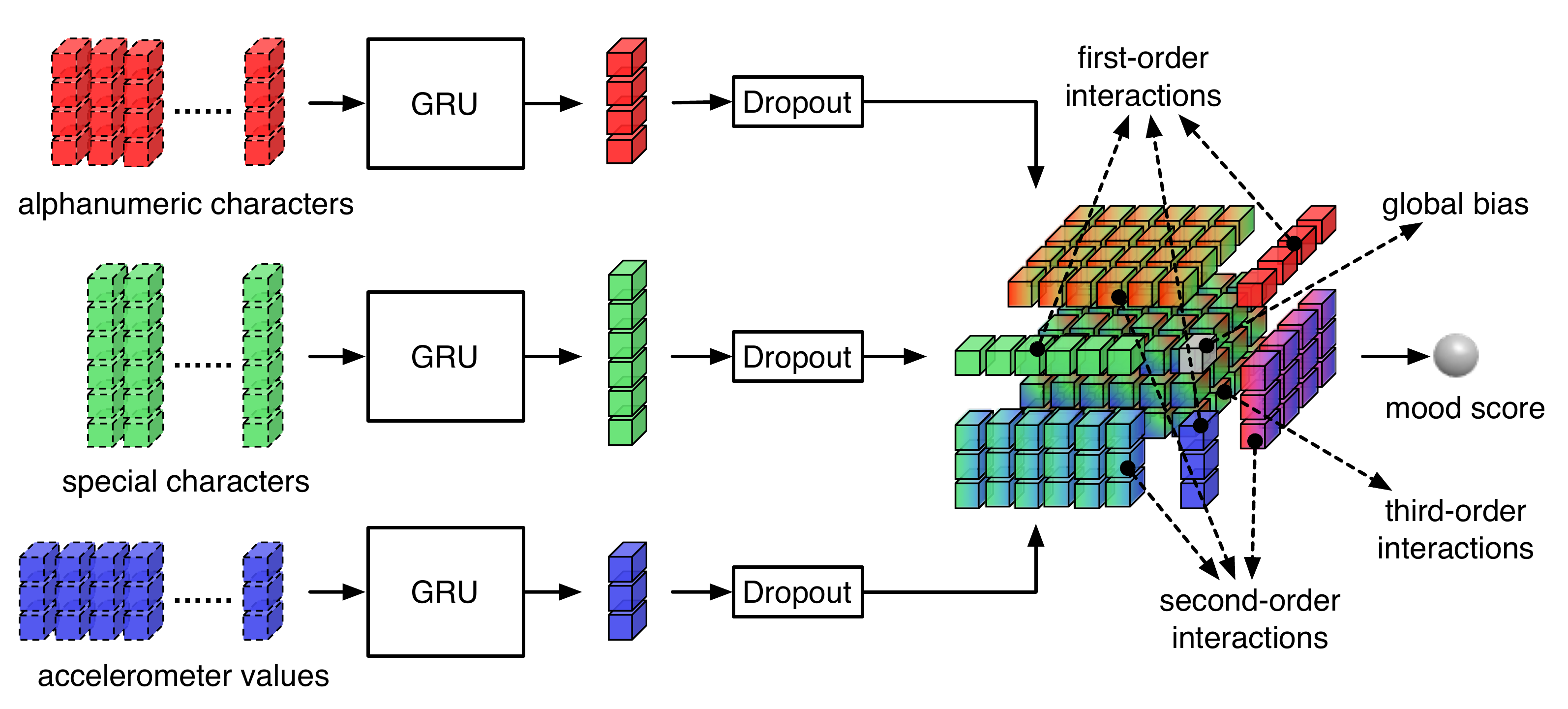}
\end{minipage}
\caption{The architecture of DeepMood with a Multi-view Machine layer for data fusion.}
\label{fig:architecture}
\end{figure*}

We study the mobile phone typing dynamics metadata on a session-level. A session is defined as beginning with a keypress which occurs after 5 or more seconds have elapsed since the last keypress and continuing until 5 or more seconds elapse between keypresses\footnote{5-second is an arbitrary threshold we set which can be changed and tuned easily.}. The duration of a session is typically less than one minute. In this manner, each participant would contribute many samples, one per phone usage session, which could benefit data analysis and model training. Each session is composed of features that are represented in multiple views or modalities ({\em e.g.}, alphanumeric characters, special characters, accelerometer values), each of which has different timestamps and densities, as shown in Figure~\ref{fig:timeline}. Modeling the multi-view time series data on such a fine-grained session-level brings up several formidable challenges:
\begin{itemize}[leftmargin=*,noitemsep,topsep=0pt]
\item\textbf{Unaligned views}: An intuitive idea for fusing the multi-view time series is to align them with each unique timestamp. However, features defined in one view would be missing for data points collected in another view. For example, a data point in special characters has no {\em acceleration} in accelerometer values or {\em distance from last key} in alphanumeric characters\footnote{This is for privacy concerns, because malicious person may be able to unscramble and recover the texts using such information.}.
\item\textbf{Dominant views}: One may also attempt to do the fusion by concatenating the multi-view time series per session. However, the views usually have different densities in a session, because the metadata are collected from different sources or sensors. For example, character-related metadata collected following a person's typing behaviours are much sparser than accelerometer values collected in the background which have 16 times more data points in our dataset. Dense views could dominate a concatenated feature space and potentially override the effects of sparse but important views.
\item\textbf{View interactions}: The multi-view time series from typing dynamics contains complementary information reflecting a person's mental health. The relationship between the digital footprints and mood states can be highly nonlinear. An effective fusion strategy is needed to explore feature interactions across different views.
\end{itemize}

In this paper, we propose a deep architecture based on late fusion, named DeepMood, to model mobile phone typing dynamics, as illustrated in Figure~\ref{fig:architecture}. The contributions of this work are threefold:
\begin{itemize}[leftmargin=*,noitemsep,topsep=0pt]
\item\textbf{Data analysis (Section~\ref{sec:data})}: We obtain interesting insights related to the digital footprints on mobile phones by analyzing the correlation between patterns of typing dynamics metadata and mood in bipolar affective disorder.
\item\textbf{A novel fusion strategy in a deep framework (Section~\ref{sec:method})}: Motivated by the aforementioned challenges that early fusion strategies ({\em i.e.}, aligning views with timestamps or concatenating views per session) would lead to the problems of unaligned or dominant views, we propose a two-stage late fusion approach for modeling the multi-view time series data. In the first stage, each view of the time series is separately modeled by a Recurrent Neural Network (RNN) \cite{mikolov2010recurrent,sutskever2011generating}. The multi-view metadata are then fused in the second stage by exploring interactions across the output vectors from each view, where three alternative approaches are developed following the idea of Multi-view Machines \cite{cao2016multi}, Factorization Machines \cite{rendle2012factorization}, or in a fully connected fashion.
\item\textbf{Empirical evaluations (Section~\ref{sec:exp})}: We conduct experiments showing that 90.31\% prediction accuracy on the depression score can be achieved based on session-level typing dynamics which reveals the potential of using mobile phone metadata to predict mood disturbance and severity. Our code is open-sourced at \textbf{\url{https://www.cs.uic.edu/~bcao1/code/DeepMood.py}}.
\end{itemize}

\section{Data}
\label{sec:data}

The data used in this work were collected from the BiAffect\footnote{\url{http://www.biaffect.com}} study which is the winner of the Mood Challenge for ResearchKit\footnote{\url{http://www.moodchallenge.com}}. During a preliminary data collection phase, for a period of 8 weeks, 40 individuals were provided a Galaxy Note 4 mobile phone which they were instructed to use as their primary phone during the study. This phone was loaded with a custom keyboard that replaced the standard Android OS keyboard. The keyboard collected metadata consisting of keypress entry time and accelerometer movement and uploaded them to the study server. In order to protect participants' privacy, individual character data with the exceptions of the backspace key and space bar were not collected.

In this work, we study the collected metadata for participants including bipolar subjects and normal controls who had provided at least one week of metadata. There are 7 participants with {\em bipolar I} disorder that involves periods of severe mood episodes from mania to depression, 5 participants with {\em bipolar II} disorder which is a milder form of mood elevation, involving milder episodes of hypomania that alternate with periods of severe depression, and 8 participants with no diagnosis per DSM-IV TR criteria \cite{kessler2005lifetime}.

Participants were administered the Hamilton Depression Rating Scale (HDRS) \cite{williams1988structured} and Young Mania Rating Scale (YMRS) \cite{young1978rating} once a week which are used as the golden standard to assess the level of depressive and manic symptoms in bipolar disorder. However, the use of these clinical rating scales requires a face-to-face patient-clinician encounter, and the level of affective symptoms is assessed during a clinical evaluation. Study findings may be unreliable when using rating scales as outcome measures due to methodological issues such as unblinding of raters and patients, differences in rater experiences and missing visits for outcome assessments \cite{demitrack1998problem,psaty2010minimizing,faurholt2016behavioral}. Thus, it motivates us to explore more objective methods with real-time data for assessing affective symptoms.

\subsection{Alphanumeric Characters}

\begin{figure}[t]
\centering
\subfigure[]{
\begin{minipage}[l]{0.45\columnwidth}
\centering
\includegraphics[width=1\textwidth]{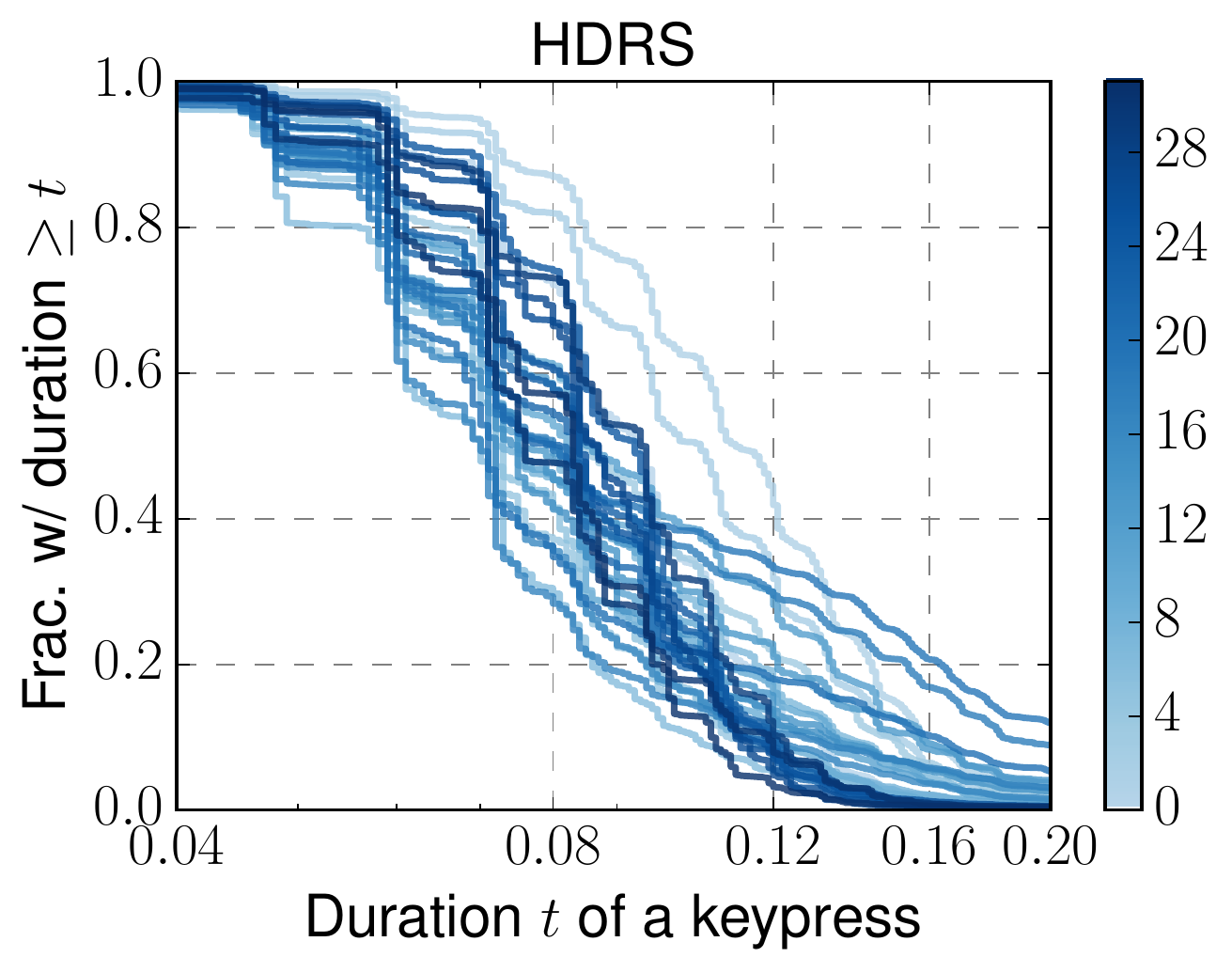}
\end{minipage}
\label{fig:cdf_dr_hdrs}
}
\subfigure[]{
\begin{minipage}[l]{0.45\columnwidth}
\centering
\includegraphics[width=1\textwidth]{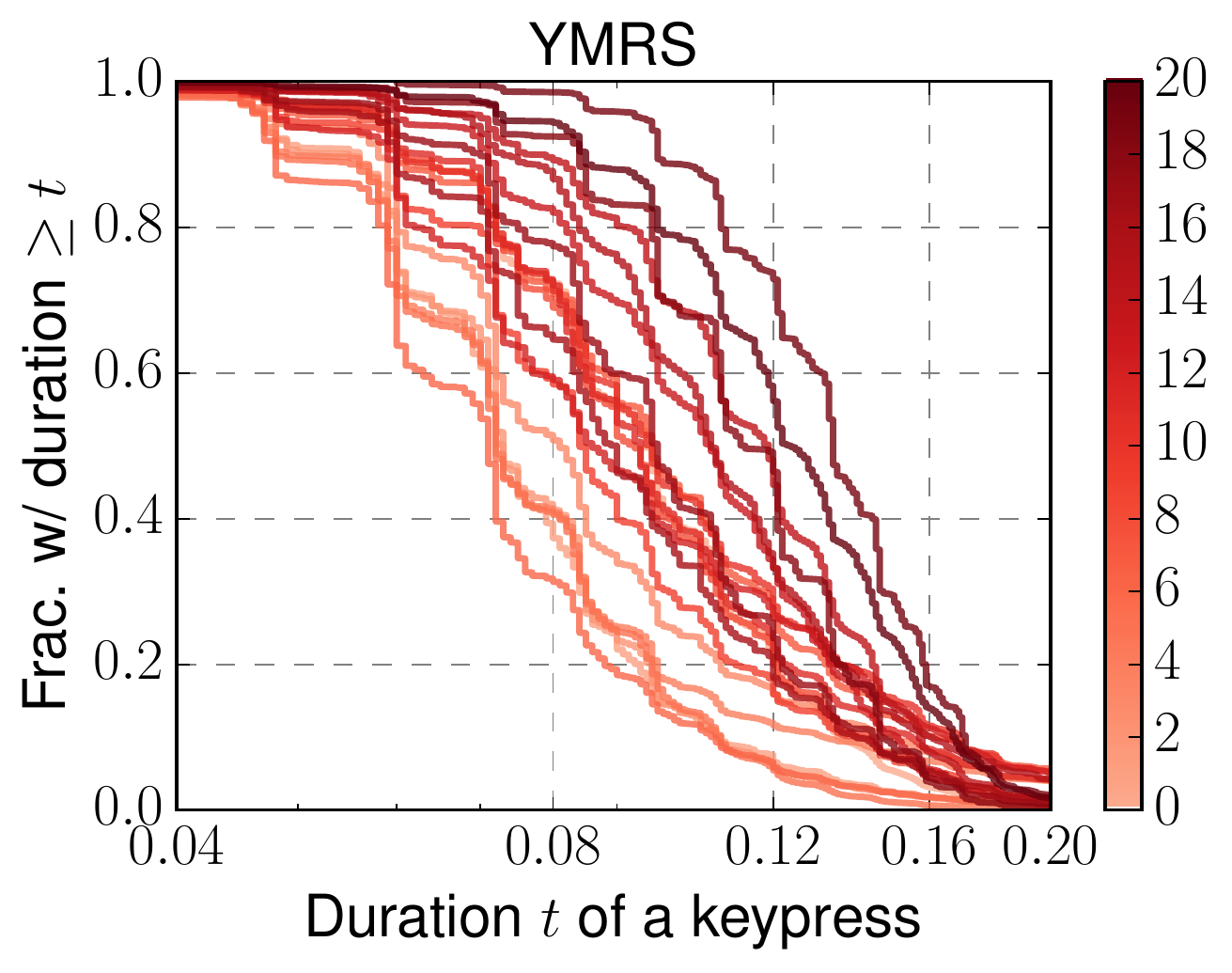}
\end{minipage}
\label{fig:cdf_dr_ymrs}
}
\caption{CCDFs of duration of a keypress.}
\label{fig:cdf_dr}
\end{figure}

\begin{figure}[t]
\centering
\subfigure[]{
\begin{minipage}[l]{0.45\columnwidth}
\centering
\includegraphics[width=1\textwidth]{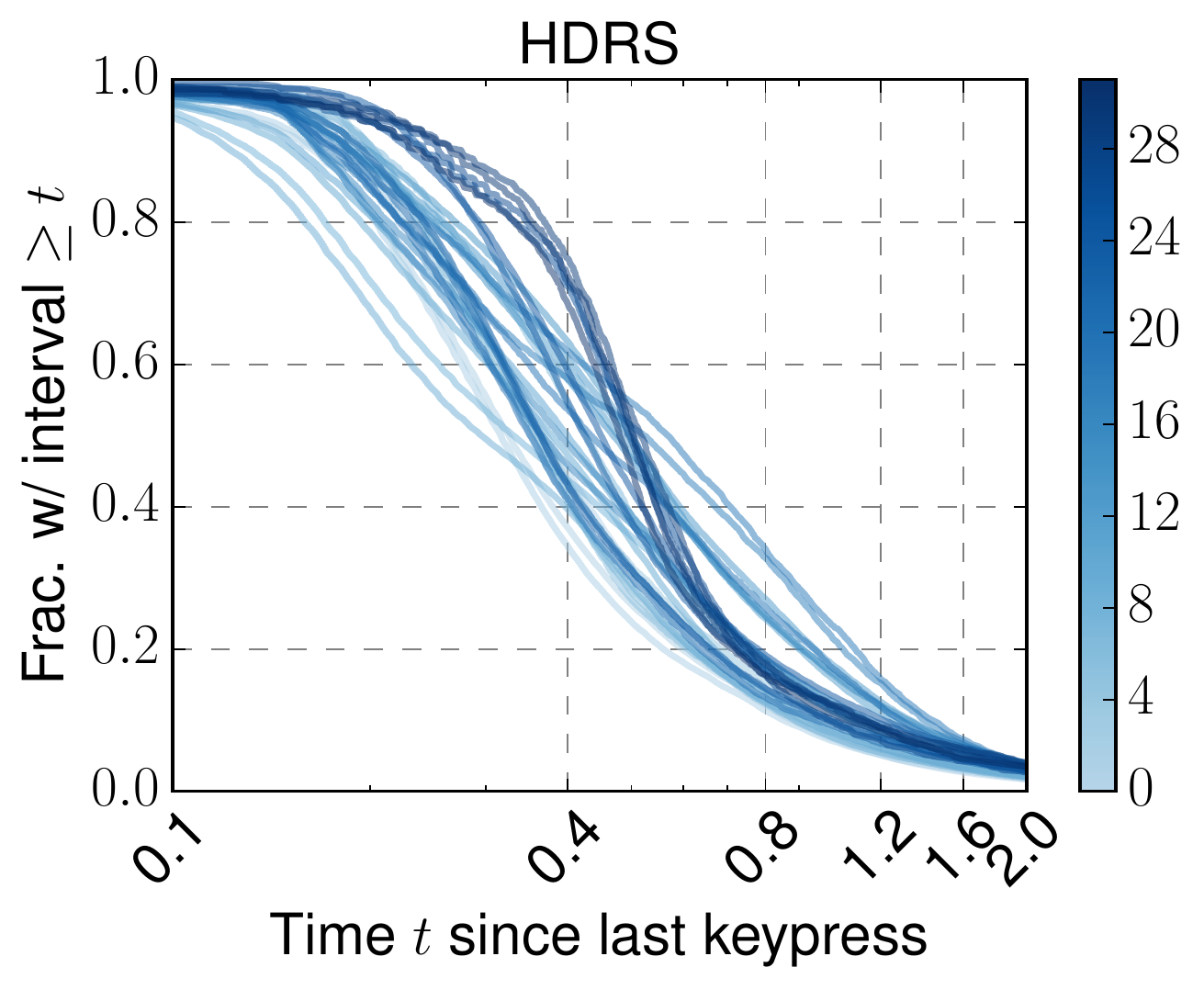}
\end{minipage}
\label{fig:cdf_dt_hdrs}
}
\subfigure[]{
\begin{minipage}[l]{0.45\columnwidth}
\centering
\includegraphics[width=1\textwidth]{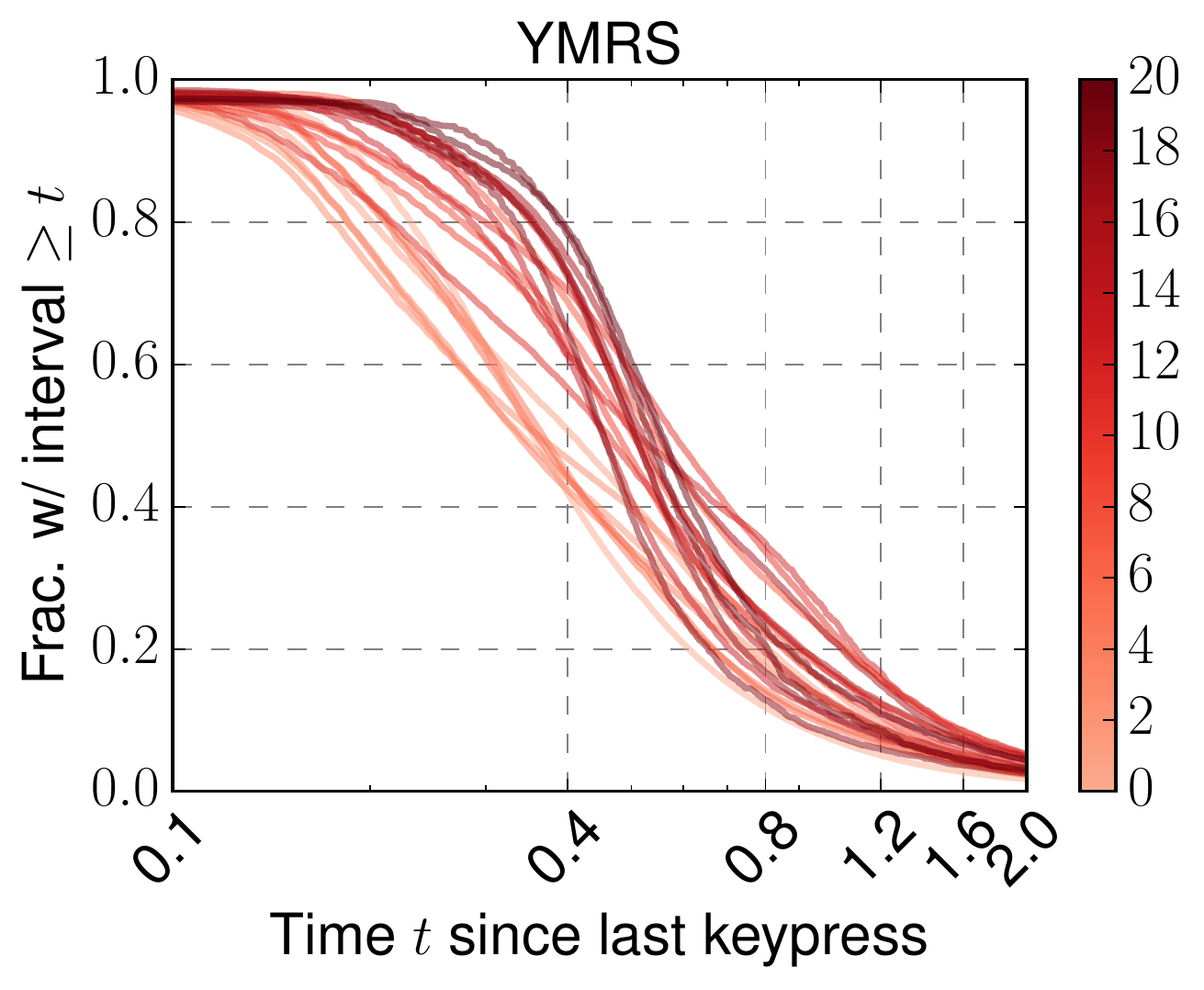}
\end{minipage}
\label{fig:cdf_dt_ymrs}
}
\caption{CCDFs of time since last keypress.}
\label{fig:cdf_dt}
\end{figure}

Due to privacy reasons, we only collected metadata for keypresses on alphanumeric characters, including duration of a keypress, time since last keypress, and distance from last key along two axises. Firstly, we aim to assess the correlation between duration of a keypress and mood states. The complementary cumulative distribution functions (CCDFs) of {\em duration of a keypress} are displayed in Figure~\ref{fig:cdf_dr}. Data points with different scores are colored differently, and the range of mood scores corresponds to the colorbar. In general, the higher the score, the darker the color and the more severe the depressive or manic symptoms. According to the Kolmogorov-Smirnov test on two samples, for all the pairs of distributions, we can reject the null hypothesis that two samples are drawn from the same distribution with significance level $\alpha=0.01$. As expected, we are dealing with a heavy-tailed distribution: (1) most keypresses are very fast with median 85ms, (2) but a non-negligible number have longer duration with 5\% using more than 155ms. Interestingly, samples with mild depression tend to have shorter duration than normal ones, while those with severe depression stand in the middle. Samples in manic symptoms seem to hold a key longer than normal ones.

Next we ask how the time since last keypress correlates with mood states. We show the CCDFs of {\em time since last keypress} in Figure~\ref{fig:cdf_dt}. Based on the Kolmogorov-Smirnov test, for 98.06\% in HDRS and 99.52\% in YMRS of the distribution pairs, we can reject the null hypothesis that two samples are drawn from the same distribution with significance level $\alpha=0.01$. Not surprisingly, this distribution is heavily skewed, with most time intervals being very short with median 380ms. However, there is a significant fraction of keypresses with much longer intervals where 5\% have more than 1.422s. We can observe that the values of time since last keypress from the normal group (with light blue/red) approximate a uniform distribution on the log scale in the range from 0.1s to 2.0s. On the contrary, this metric from samples with mood disturbance (with dark blue/red) shows a more skewed distribution with a few values on the two tails and majority centered between 0.4s and 0.8s. In other words, healthy people show a good range of reactivity that gets lost in mood disturbance where the range is more restricted.

\begin{figure}[t]
\centering
\subfigure[]{
\begin{minipage}[l]{0.45\columnwidth}
\centering
\includegraphics[width=1\textwidth]{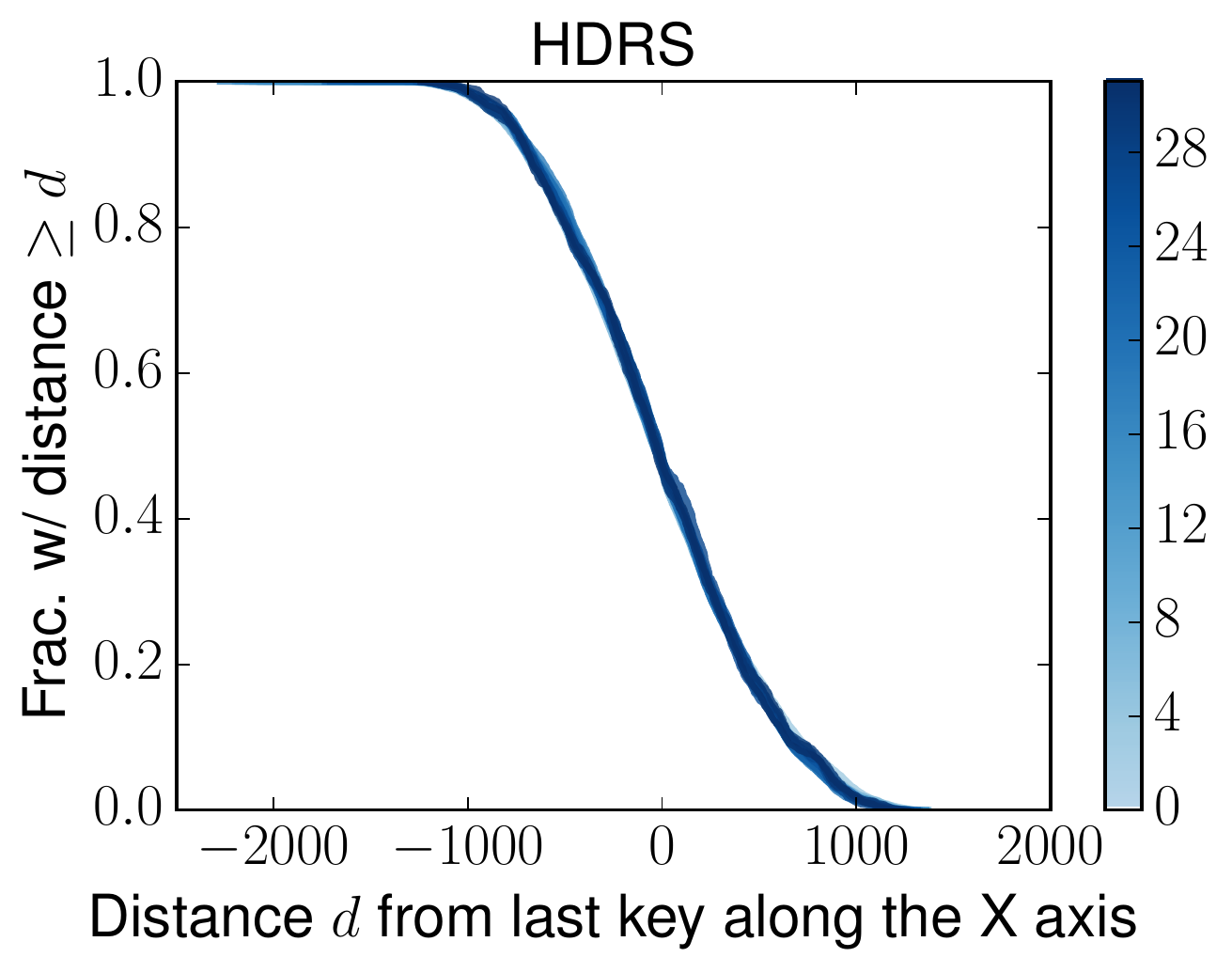}
\end{minipage}
}
\subfigure[]{
\begin{minipage}[l]{0.45\columnwidth}
\centering
\includegraphics[width=1\textwidth]{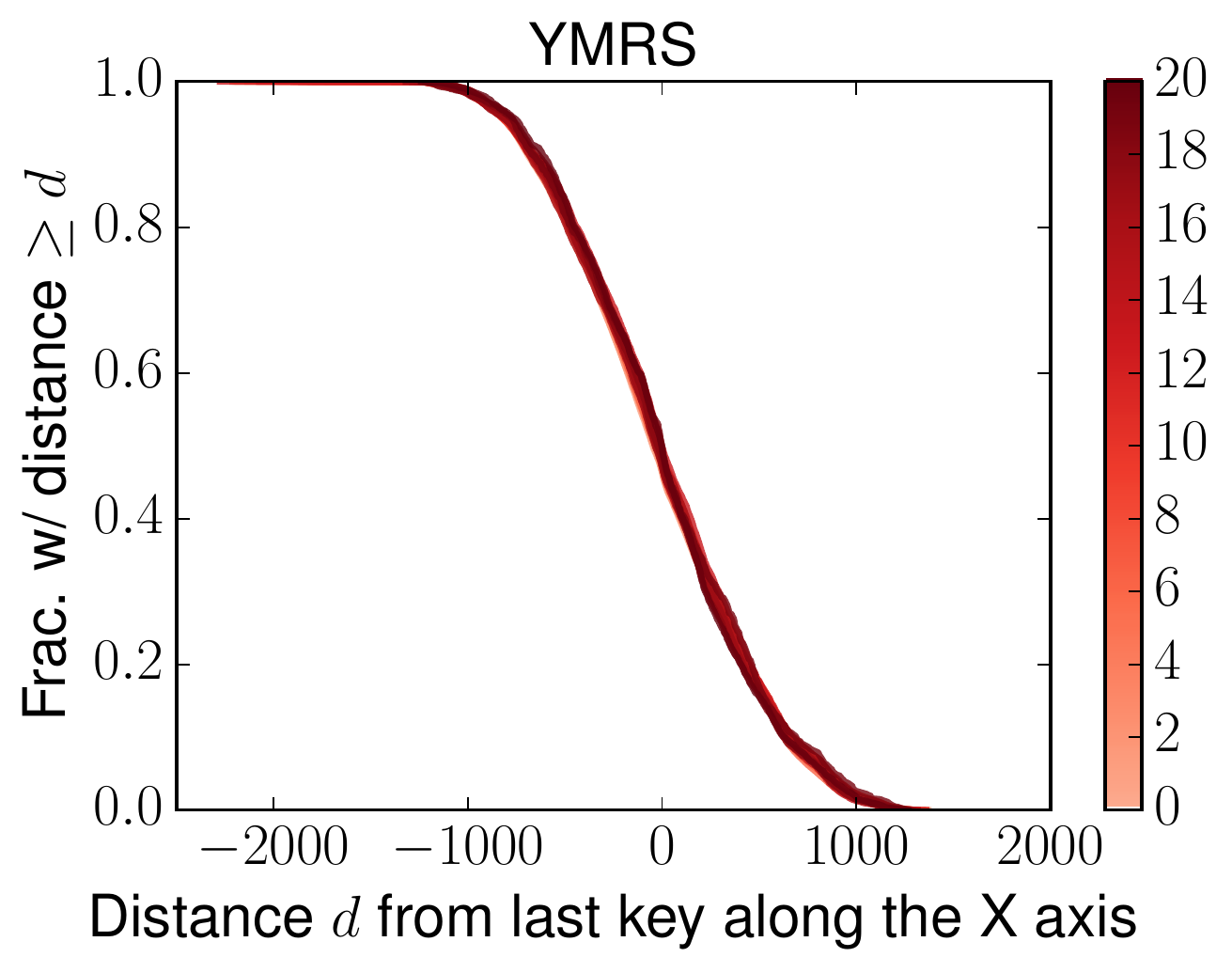}
\end{minipage}
}
\subfigure[]{
\begin{minipage}[l]{0.45\columnwidth}
\centering
\includegraphics[width=1\textwidth]{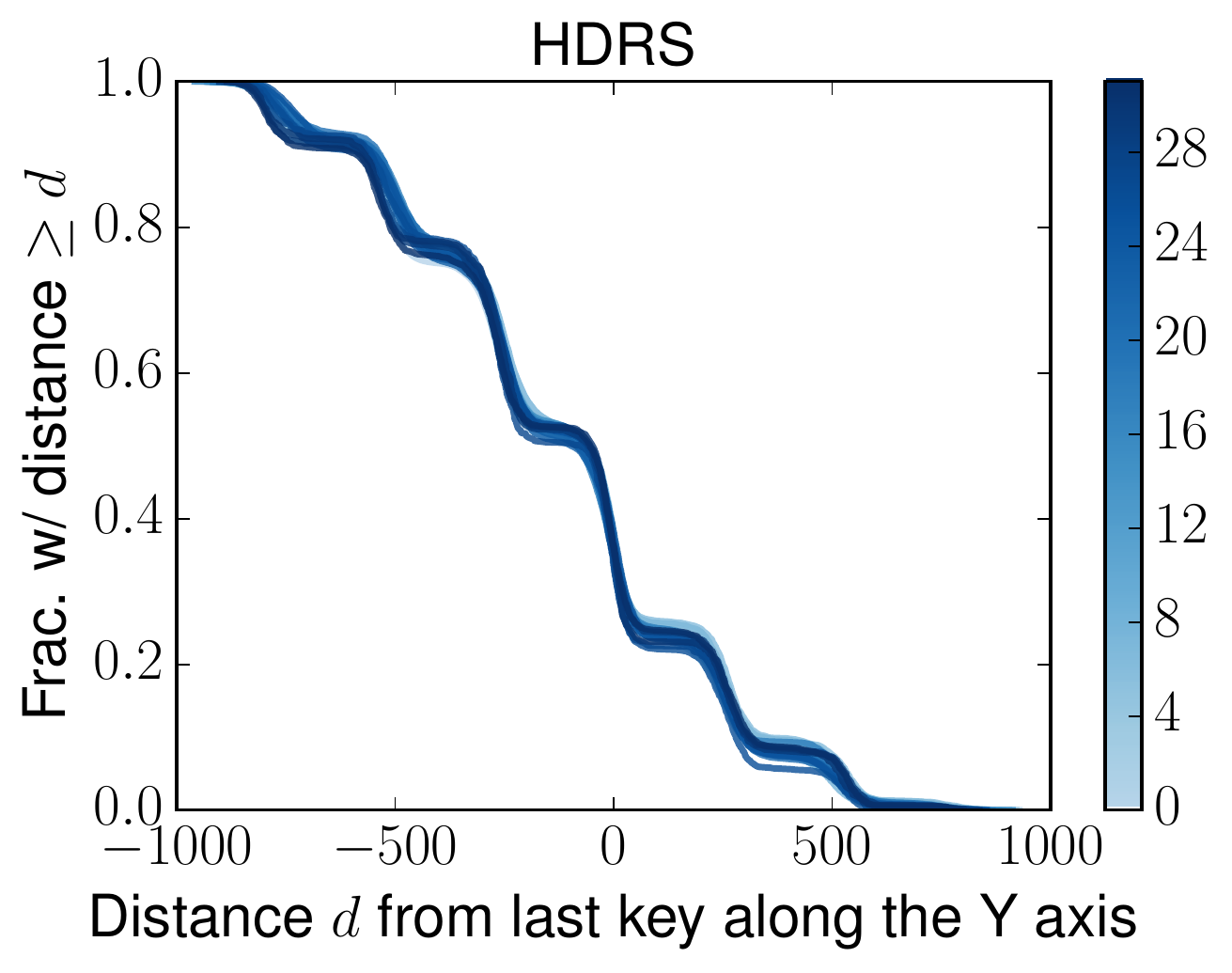}
\end{minipage}
}
\subfigure[]{
\begin{minipage}[l]{0.45\columnwidth}
\centering
\includegraphics[width=1\textwidth]{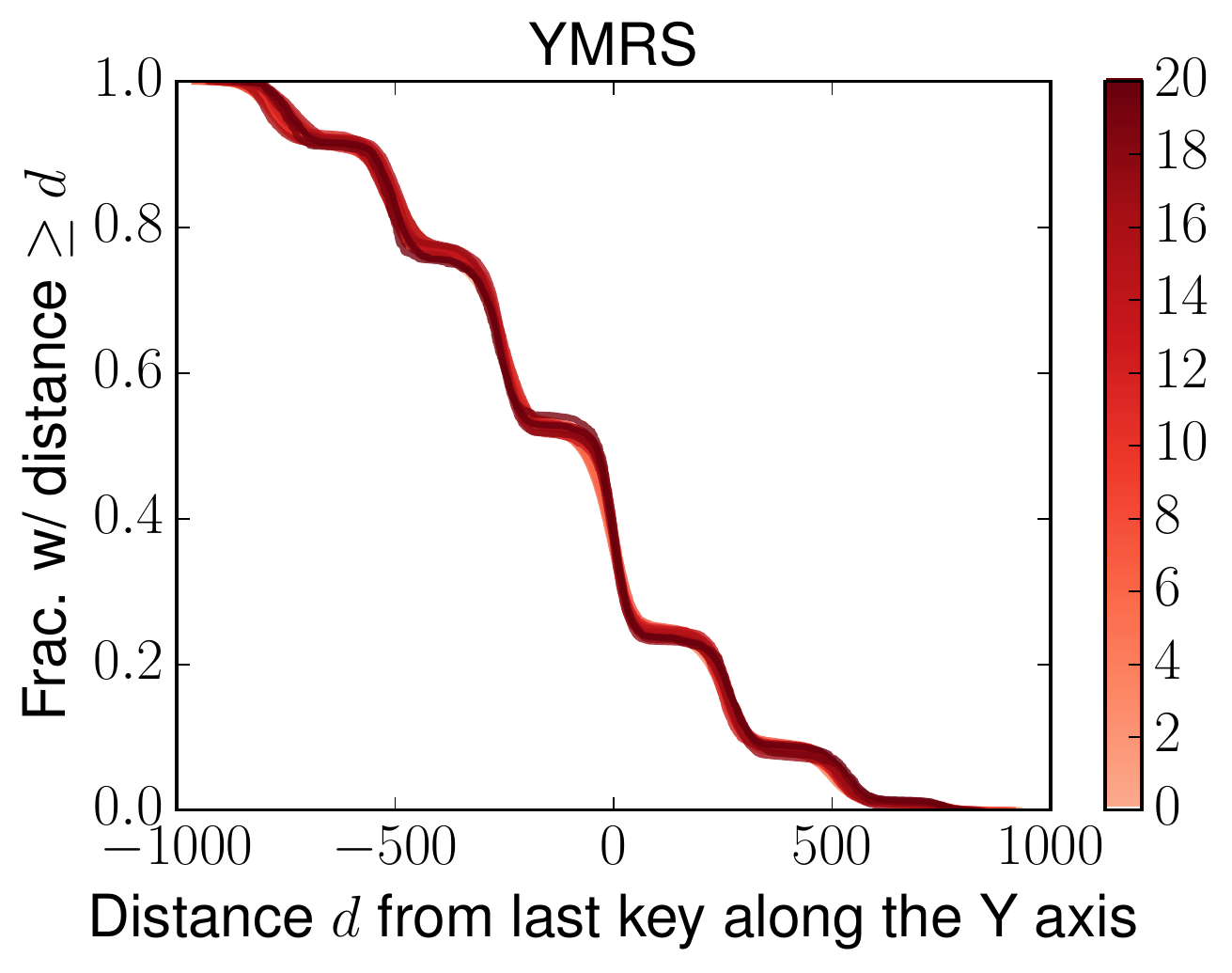}
\end{minipage}
}
\caption{CCDFs of distance from last key along two axises. Note that lines are almost identical.}
\label{fig:cdf_dx_dy}
\end{figure}

Figure~\ref{fig:cdf_dx_dy} shows the CCDFs of {\em distance from last key} along two axises which can be considered as a sort of very rough proxy of the semantic content of people's typing.
No distinction can be observed across different mood states, because there are no dramatic differences in the manner in which depressive or manic people type compared to controls.


\begin{figure}[t]
\centering
\begin{minipage}[l]{\columnwidth}
\centering
\includegraphics[width=1\textwidth]{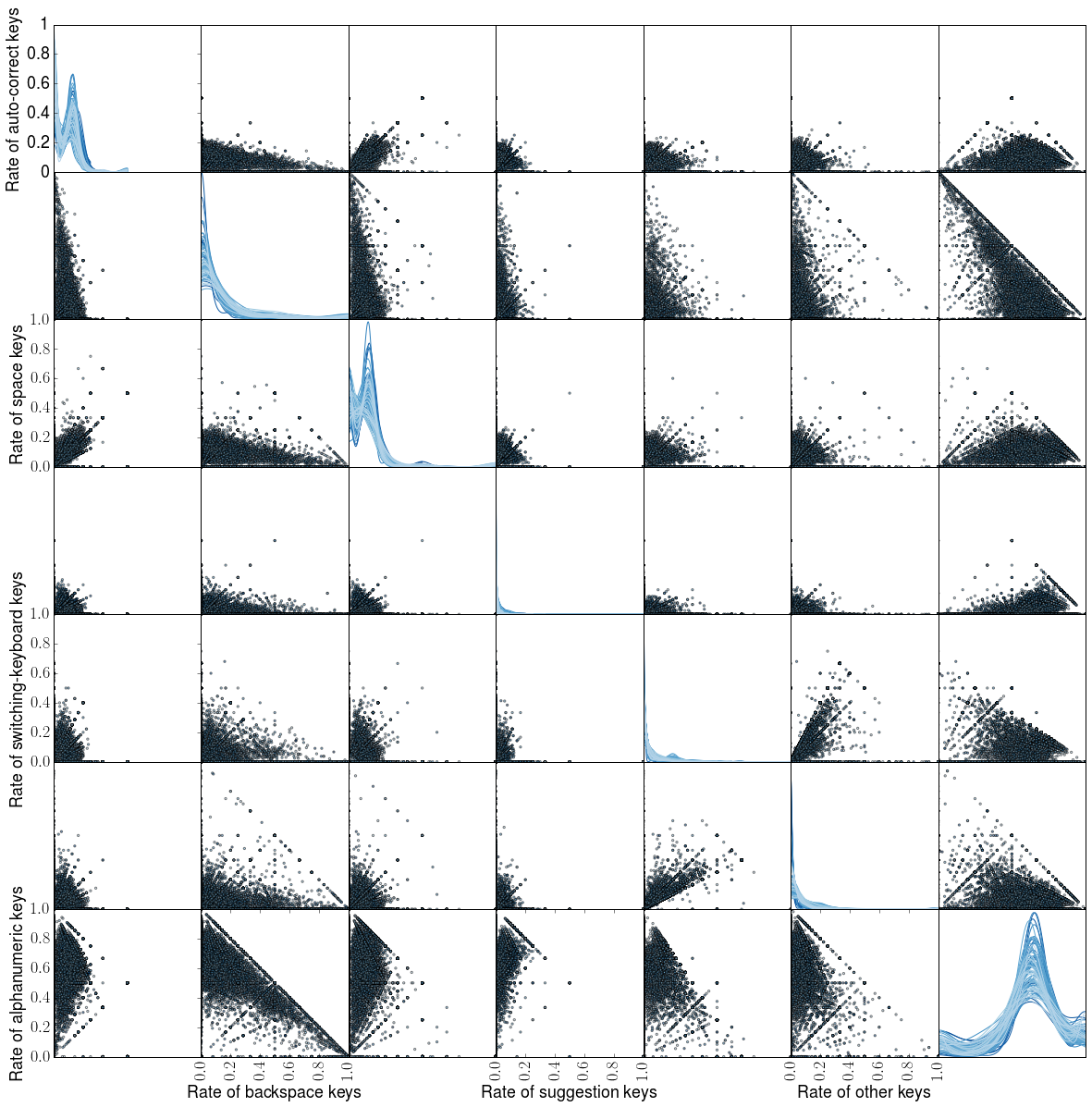}
\end{minipage}
\caption{Scatter plot between rates of different keys.}
\label{fig:scatter}
\end{figure}

\subsection{Special Characters}

In this view, we use one-hot-encoding for typing behaviors other than alphanumeric characters, including {\em auto-correct}, {\em backspace}, {\em space}, {\em suggestion}, {\em switching-keyboard} and {\em other}. They are usually sparser than alphanumeric characters. Figure~\ref{fig:scatter} shows the scatter plot between rates of these special characters as well as alphanumeric ones in a session where the color of a dot/line corresponds to the HDRS score. Although no obvious distinction can be found between mood states, we can observe some interesting patterns: the rate of alphanumeric keys is negatively correlated with the rate of {\em backspace} (from the subfigure at the 2nd row, 7th column), while the rate of {\em switching-keyboard} is positively correlated with the rate of {\em other keys} (from the subfigure at the 5th row, 6th column). On the diagonal there are kernel density estimations. It shows that the rate of alphanumeric characters is generally high in a session, followed by {\em auto-correct}, {\em space}, {\em backspace}, {\em etc.} Similar patterns can be found from the plot of YMRS which is omitted here.

\begin{figure}[t]
\centering
\subfigure[]{
\begin{minipage}[l]{0.45\columnwidth}
\centering
\includegraphics[width=1\textwidth]{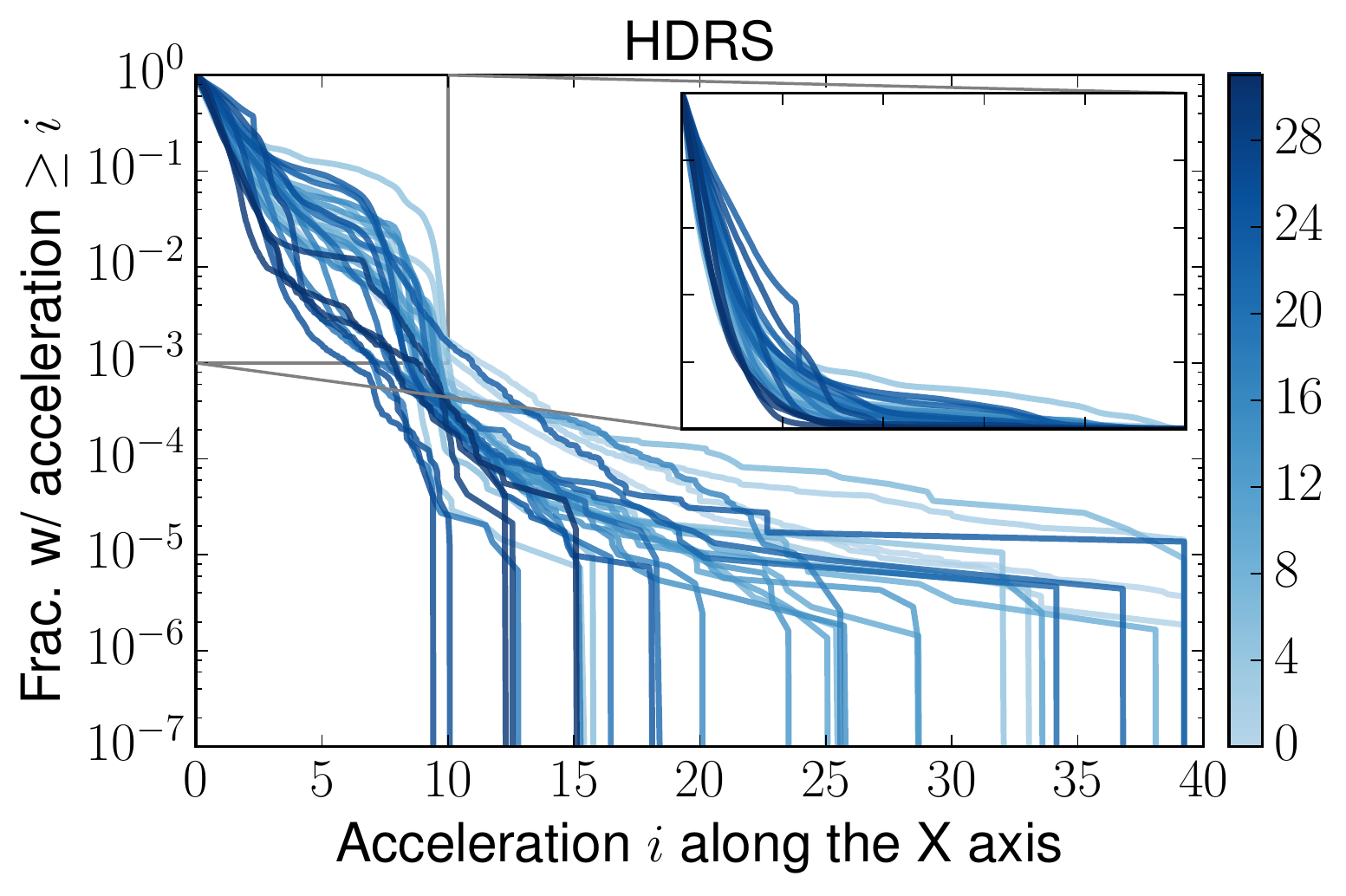}
\end{minipage}
}
\subfigure[]{
\begin{minipage}[l]{0.45\columnwidth}
\centering
\includegraphics[width=1\textwidth]{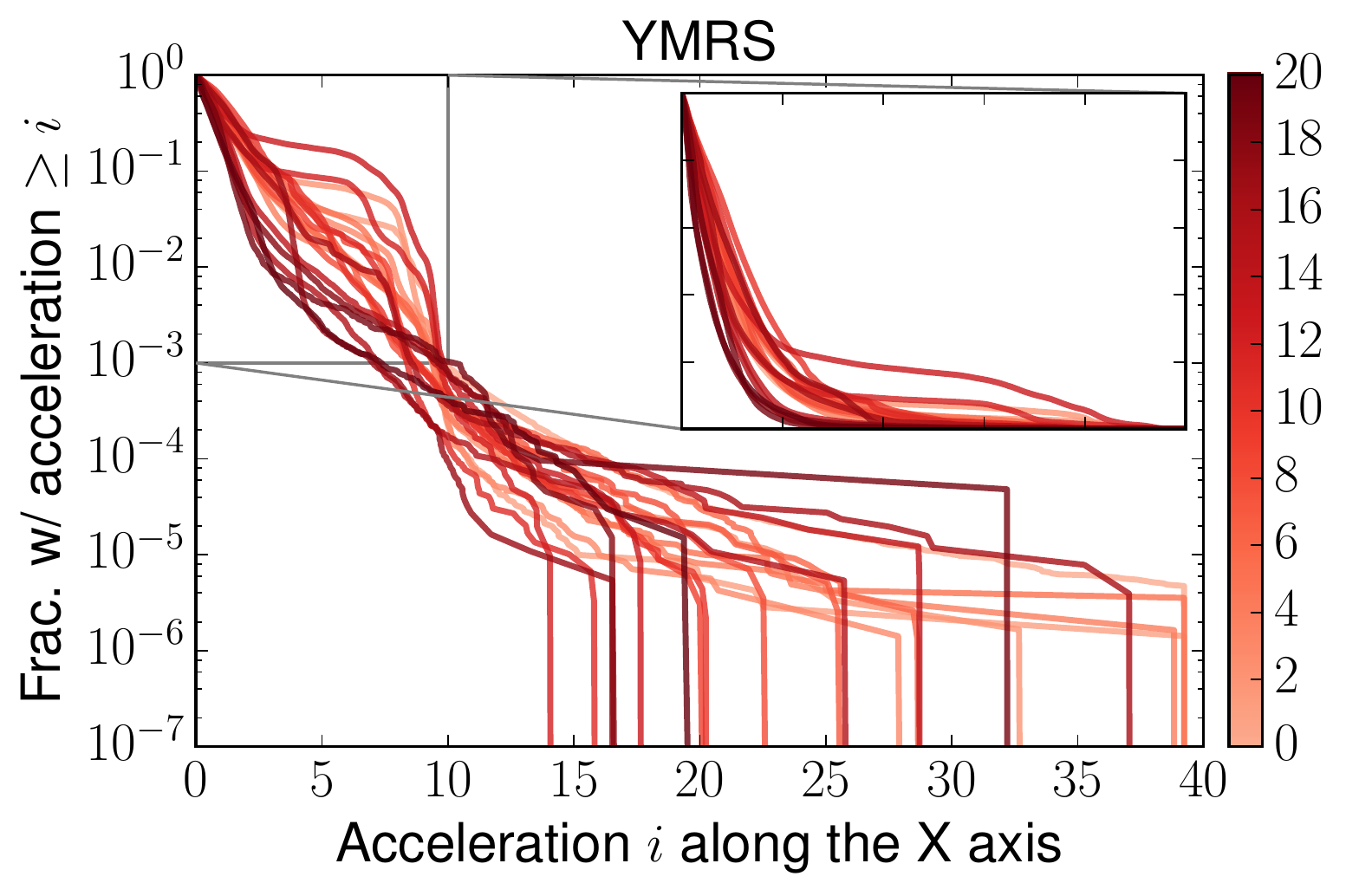}
\end{minipage}
}
\subfigure[]{
\begin{minipage}[l]{0.45\columnwidth}
\centering
\includegraphics[width=1\textwidth]{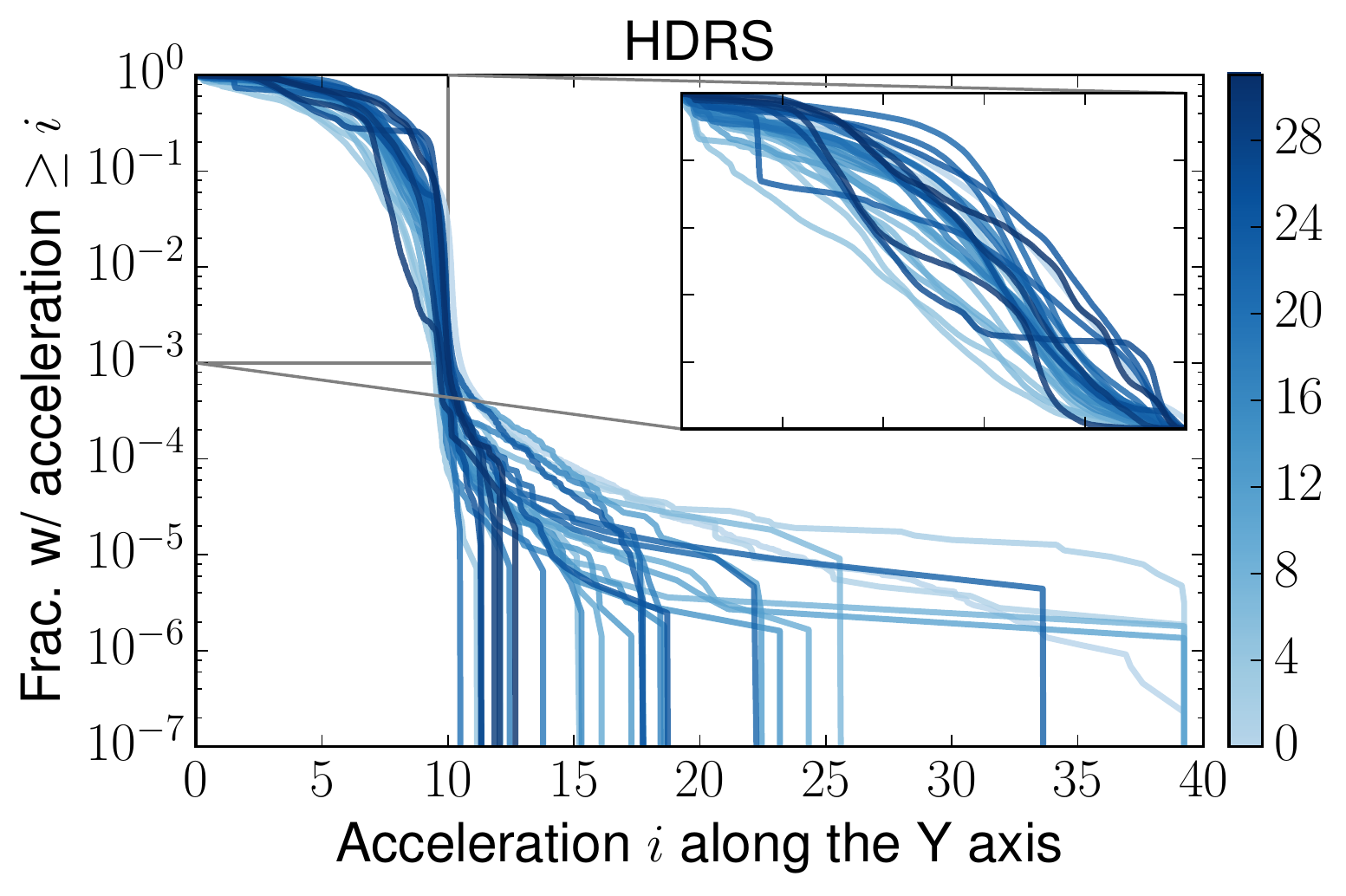}
\end{minipage}
}
\subfigure[]{
\begin{minipage}[l]{0.45\columnwidth}
\centering
\includegraphics[width=1\textwidth]{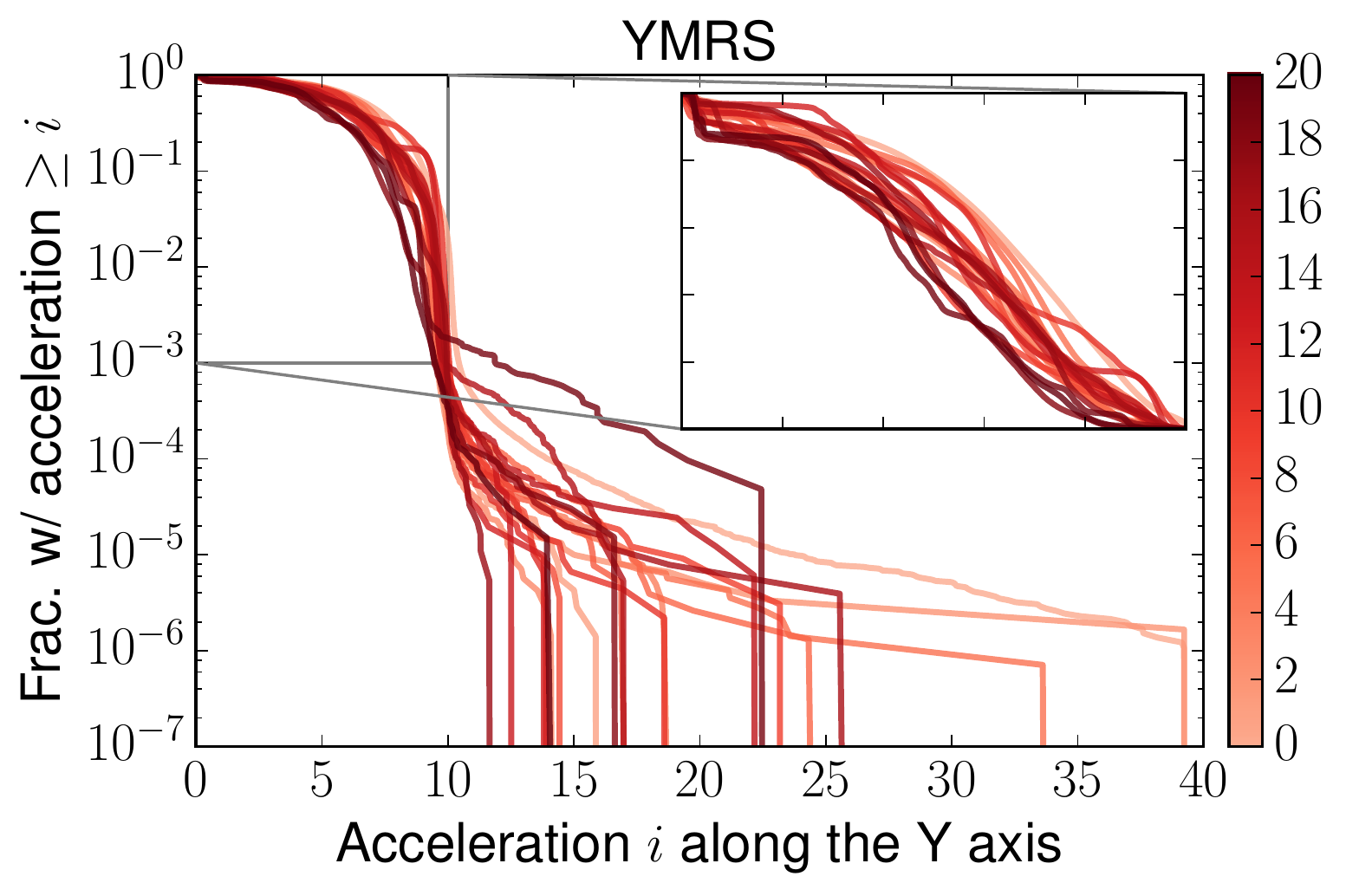}
\end{minipage}
}
\subfigure[]{
\begin{minipage}[l]{0.45\columnwidth}
\centering
\includegraphics[width=1\textwidth]{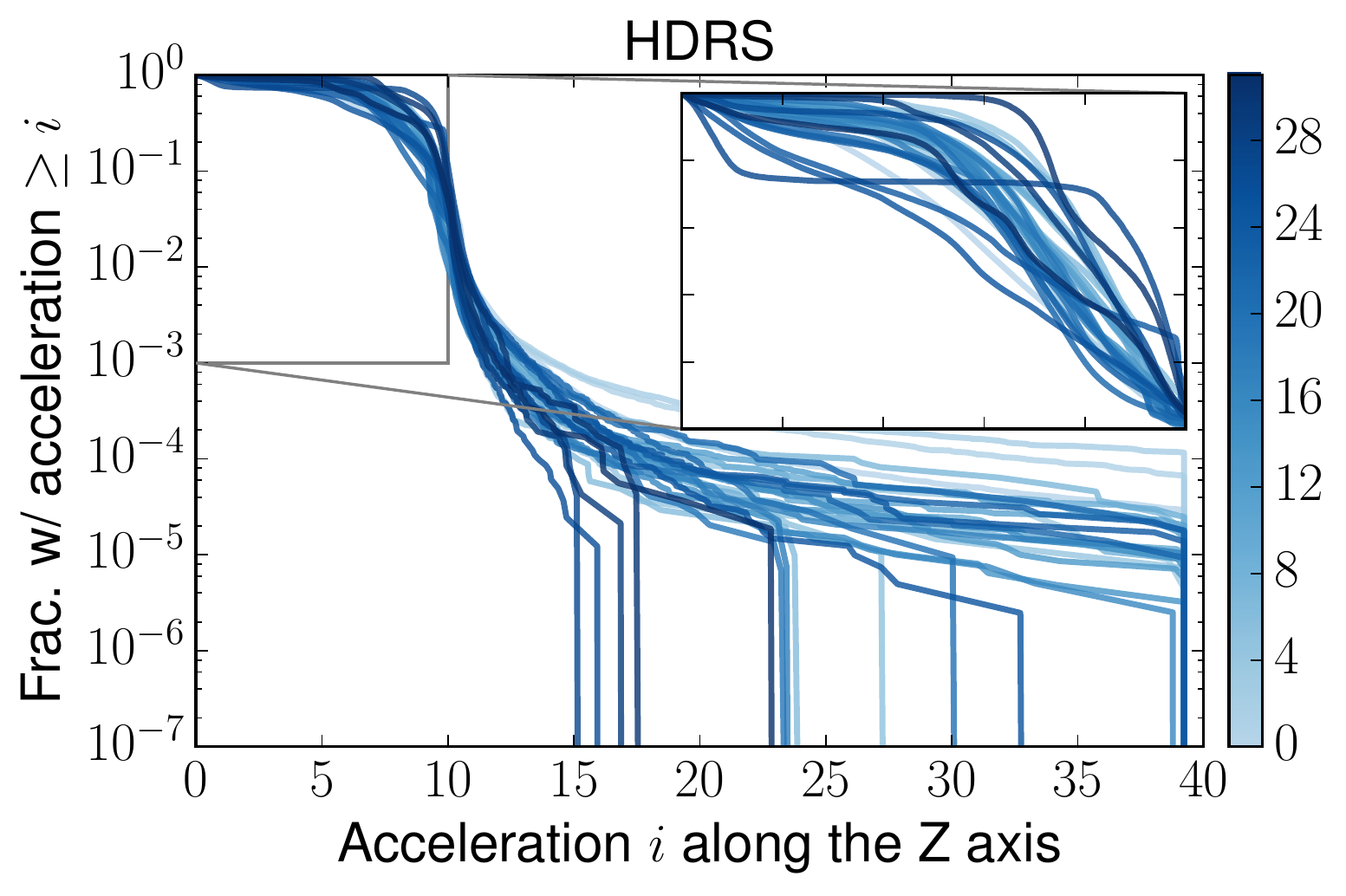}
\end{minipage}
}
\subfigure[]{
\begin{minipage}[l]{0.45\columnwidth}
\centering
\includegraphics[width=1\textwidth]{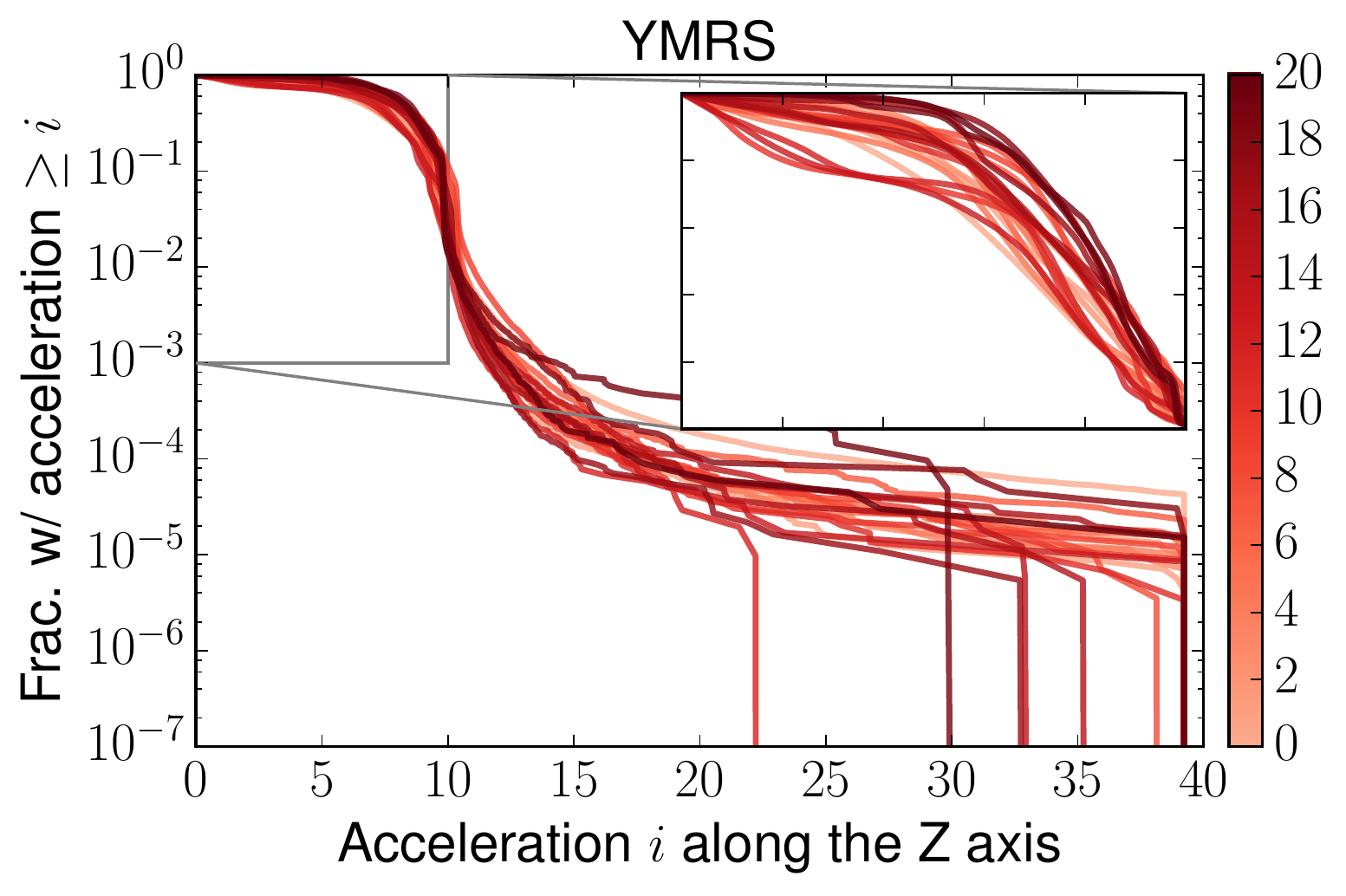}
\end{minipage}
}
\caption{CCDFs of absolute acceleration along three axises.}
\label{fig:cdf_accel}
\end{figure}

\subsection{Accelerometer Values}

Accelerometer values are recorded every 60ms in the background during an active session regardless of a person's typing speed, thereby making them much denser than alphanumeric characters. The CCDFs of absolute accelerometer values along three axises are displayed in Figure~\ref{fig:cdf_accel}. Data points with different mood scores are colored differently, and the higher the score, the more severe the depressive or manic symptoms. According to the Kolmogorov-Smirnov test on two samples, for all the pairs of distributions, we can reject the null hypothesis that two samples are drawn from the same distribution with significance level $\alpha=0.01$. Note that the vertical axis of the non-zoomed plots is on a log scale. We observe a heavy-tailed distribution for all three axises and for both HDRS and YMRS, with more than 99\% of data points being less than 7.45, 9.97 and 10.56 along X, Y and Z axis, respectively. By zooming into data points at the ``head'' of the distribution on a regular scale, we can see different patterns on the absolute acceleration along different axises. There is a nearly uniform distribution of absolute acceleration along the Y axis in the range from 0 to 10, while the majority along the X axis lie between 0 and 2, and the majority along the Z axis lie between 6 and 10. An interesting observation is that compared with normal ones, samples with mood disturbance tend to have larger accelerations along the Z axis, and smaller accelerations along the Y axis. Hence, we suspect that people in a normal mood state prefer to hold their phone towards to themselves, while people in depressive or manic symptoms are more likely to lay their phone with an angle towards to the horizon, given that data were collected only when the phone was in a portrait position.




See Table~\ref{tab:stats} for more information about the statistics of the dataset. Note that the length of a sequence is measured in terms of the number of data points in a sample rather than the duration in time.

\begin{table}[t]
\centering
\caption{Statistics of the dataset.}
\label{tab:stats}
\begin{tabular}{||l|c|c|c||}
\hline
Statistics & {\ch} & {\nonch} & {\accel} \\
\hline\hline
\# data points      & 836,027 & 538,520 & 14,237,503 \\
\# sessions         & 34,993 & 33,385 & 37,647 \\
mean length         & 24 & 16 & 378 \\
median length       & 14 & 9 & 259 \\
maximum length      & 538 & 437 & 90,193 \\
\hline
\end{tabular}
\end{table}

\section{DeepMood Architecture based on Late Fusion}
\label{sec:method}

In this paper, we propose an end-to-end deep architecture, named DeepMood, to model mobile phone typing dynamics. Specifically, DeepMood provides a late fusion framework. It first models each view of the time series data separately using Gated Recurrent Unit (GRU) \cite{cho2014learning}, a simplified version of Long Short-Term Memory (LSTM) \cite{hochreiter1997long}. It then fuses the output of the GRU from each view. As the GRU extracts a latent feature representation out of each time series, where the notions of sequence length and sampling time points are removed from the latent space, this avoids the problem of dealing directly with the heterogeneity of the time series from each view. Following the idea of Multi-view Machines \cite{cao2016multi}, Factorization Machines \cite{rendle2012factorization}, or in a conventional fully connected fashion, three alternative fusion layers are designed to integrate the complementary information in the multi-view time series to produce a prediction on the mood score. The architecture is illustrated in Figure~\ref{fig:architecture}.

\subsection{Modeling One View}

Each view in the metadata is essentially a time series whose length can vary a lot across sessions that largely depends on the duration of a session. In order to model the dynamic sequential correlations in each time series, we adopt the RNN architecture \cite{mikolov2010recurrent,sutskever2011generating} which keeps hidden states over a sequence of elements and updates the hidden state $\mathbf{h}_k$ by the current input $\mathbf{x}_k$ as well as the previous hidden state $\mathbf{h}_{k-1}$ where $k>1$ with a recurrent function:
\begin{equation}
\mathbf{h}_k = f(\mathbf{x}_k, \mathbf{h}_{k-1})
\end{equation}

The simplest form of an $\operatorname{RNN}$ is as follows:
\begin{equation}
\mathbf{h}_k = \sigma(\mathbf{W}\mathbf{x}_k+\mathbf{U}\mathbf{h}_{k-1})
\end{equation}
where $\mathbf{W} \in \mathbb{R}^{d_h \times d_x}, \mathbf{U} \in \mathbb{R}^{d_h \times d_h}$ are model parameters that need to be learned, $d_x$ and $d_h$ are the input dimension and the number of recurrent units, respectively. $\sigma(\cdot)$ is a nonlinear transformation function such as tanh, sigmoid, and rectified linear unit (ReLU). Since RNNs in such a form would fail to learn long term dependencies due to the exploding and the vanishing gradient problem \cite{bengio1994learning,hochreiter1998vanishing}, they are not suitable to learn dependencies from a long input sequence in practice.

To make the learning procedure more effective over long sequences, the GRU \cite{cho2014learning} is proposed as a variation of the LSTM unit \cite{hochreiter1997long}. The GRU has been attracting great attentions since it overcomes the vanishing gradient problem in traditional RNNs and is more efficient than the LSTM in some tasks \cite{chung2014empirical}. The GRU is designed to learn from previous timestamps with long time lags of unknown size between important timestamps via memory units that enable the network to learn to both update and forget hidden states based on new inputs.

A typical GRU is formulated as:
\begin{equation}
\begin{aligned}
\mathbf{r}_k &= \text{sigmoid} (\mathbf{W}_r\mathbf{x}_k + \mathbf{U}_r\mathbf{h}_{k - 1}) \hfill \\
\mathbf{z}_k &= \text{sigmoid} (\mathbf{W}_z\mathbf{x}_k + \mathbf{U}_z\mathbf{h}_{k - 1}) \hfill \\
\tilde{\mathbf{h}}_k &= \text{tanh} (\mathbf{W}\mathbf{x}_k + \mathbf{U}(\mathbf{r}_k \odot \mathbf{h}_{k - 1})) \hfill \\
\mathbf{h}_k &= \mathbf{z}_k \odot \mathbf{h}_{k - 1} + (1 - \mathbf{z}_k) \odot \tilde {\mathbf{h}}_k \hfill \\
\end{aligned}
\end{equation}
where $\odot$ is the element-wise multiplication operator, a reset gate $\mathbf{r}_k$ allows the GRU to forget the previously computed state $\mathbf{h}_{k-1}$, and an update gate $\mathbf{z}_k$ balances between the previous state $\mathbf{h}_{k-1}$ and the candidate state $\tilde{\mathbf{h}}_k$. The hidden state $\mathbf{h}_k$ can be considered as a compact representation of the input sequence from $\mathbf{x}_1$ to $\mathbf{x}_k$.

\subsection{Late Fusion on Multiple Views}

Here we pursue a late fusion strategy to integrate the output vectors of the GRU units on these time series data from different views. This avoids the issues of alignment and diverse frequencies among the time series under different views when performing early fusion directly on the input data.  

\begin{figure}[t]
\centering
\subfigure[Fully connected layer.]{
\begin{minipage}[l]{0.9\columnwidth}
\centering
\includegraphics[width=1\textwidth]{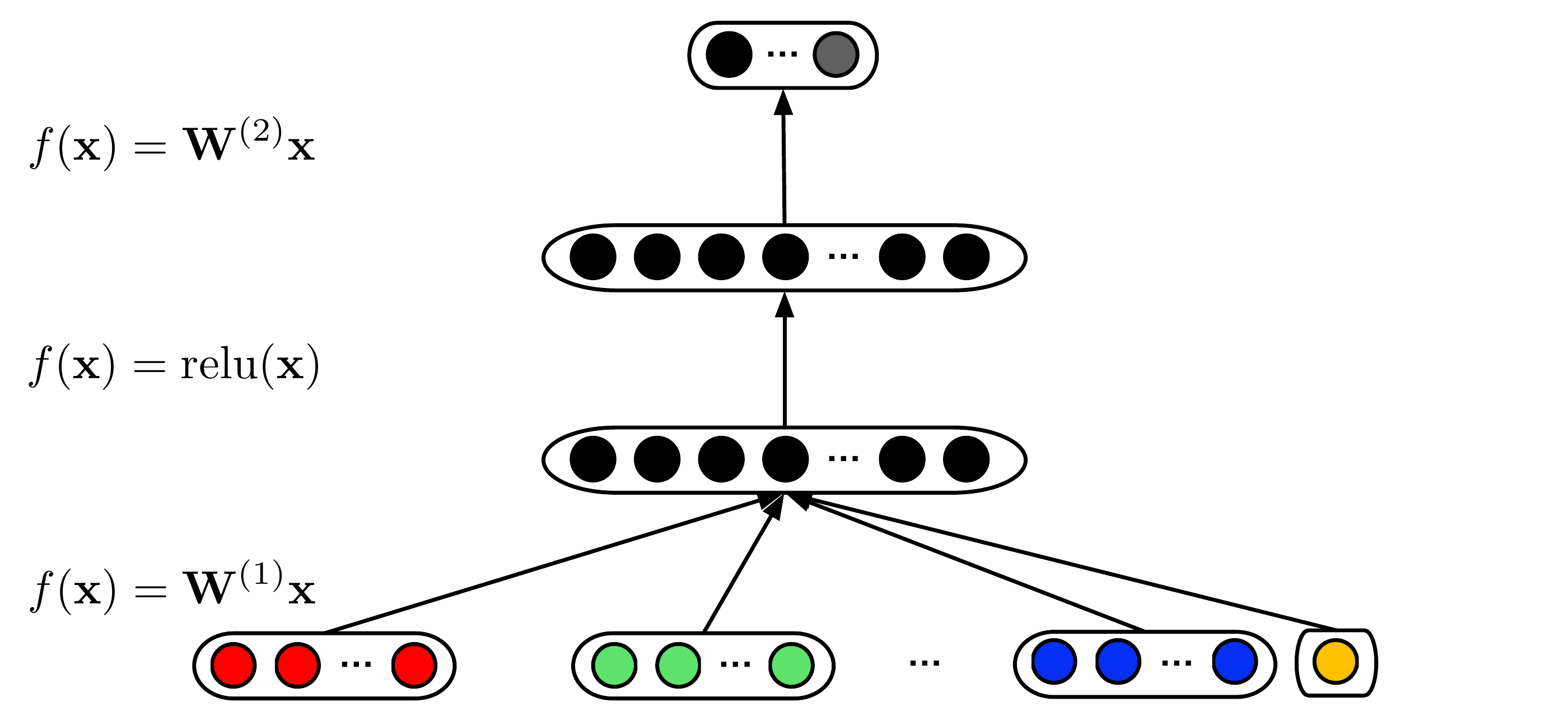}
\end{minipage}
\label{fig:fig_nn}
}
\subfigure[Factorization Machine layer.]{
\begin{minipage}[l]{0.9\columnwidth}
\centering
\includegraphics[width=1\textwidth]{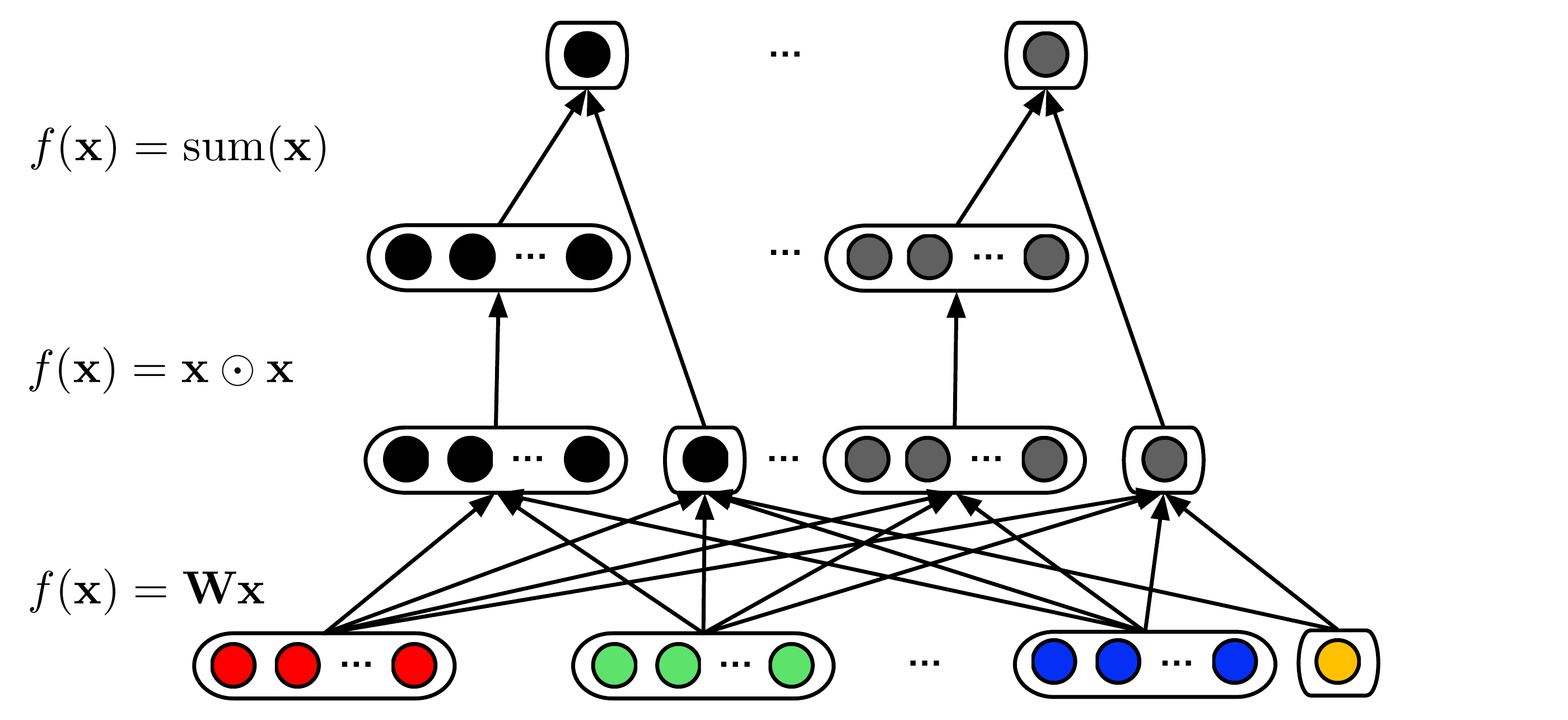}
\end{minipage}
\label{fig:fig_fm}
}
\subfigure[Multi-view Machine layer.]{
\begin{minipage}[l]{0.9\columnwidth}
\centering
\includegraphics[width=1\textwidth]{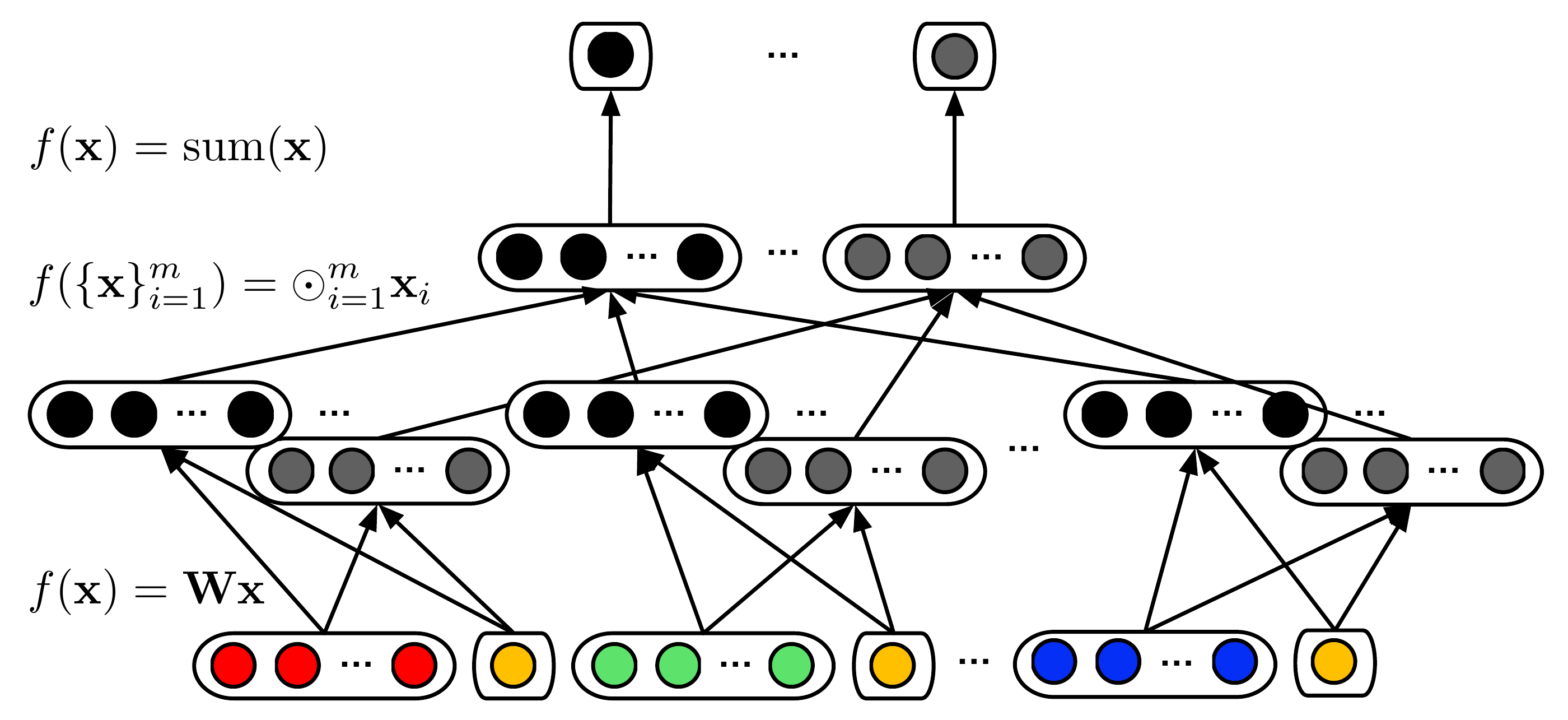}
\end{minipage}
\label{fig:fig_mvm}
}
\caption{A comparison of different strategies for fusing multi-view data from the perspective of computational graph. Red, green and blue represent features coming from different views, and yellow represents a constant signal ``1'' which models the bias.}
\label{fig:fig_graph}
\end{figure}

In the following we study alternative methods for performing late fusion. These include not only the straightforward approach based on adding a fully connected layer to concatenate the features from different views, but also novel approaches to capture interactions among the features across multiple views by exploring the concept of Factorization Machines \cite{rendle2012factorization} to capture the second-order interactions as well as the concept of Multi-view Machines \cite{cao2016multi} to capture higher order interactions as shown in Figure~\ref{fig:fig_graph}.

We denote the output vectors at the end of a sequence from the $p$-th view as $\mathbf{h}^{(p)}$. We can consider $\{\mathbf{h}^{(p)}\in\mathbb{R}^{d_h}\}_{p=1}^m$ as multi-view data where $m$ is the number of views.

\noindent\textbf{Fully connected layer}.
In order to generate a prediction on the mood score, a straightforward idea is to first concatenate features from multiple views together, {\em i.e.}, $\mathbf{h} = [\mathbf{h}^{(1)}; \mathbf{h}^{(2)}; \cdots; \mathbf{h}^{(m)}] \in \mathbb{R}^d$, where $d$ is the total number of multi-view features, and typically $d=md_h$ for one-directional RNNs and $d=2md_h$ for bidirectional RNNs. We then feed forward $\mathbf{h}$ into one or several fully connected neural network layers with a nonlinear function $\sigma(\cdot)$ in between.
\begin{equation}
\begin{aligned}
\mathbf{q} &= \text{relu}(\mathbf{W}^{(1)}[\mathbf{h}; 1]) \\
\hat{\mathbf{y}} &= \mathbf{W}^{(2)}\mathbf{q}
\end{aligned}
\end{equation}
where $\mathbf{W}^{(1)} \in \mathbb{R}^{k' \times (d+1)}, \mathbf{W}^{(2)} \in \mathbb{R}^{c \times k'}$, $k'$ is the number of hidden units, $c$ is the number of classes, and the constant signal ``1'' is to model the global bias. Note that here we consider only one hidden layer between the input layer and the final output layer as shown in Figure~\ref{fig:fig_nn}.

\noindent\textbf{Factorization Machine layer}.
Rather than capturing nonlinearity through the transformation function, we consider explicitly modeling feature interactions between input units as shown in Figure~\ref{fig:fig_fm}.
\begin{equation}
\begin{aligned}
\mathbf{q}_a &= \mathbf{U}_a\mathbf{h} \\
b_a &= \mathbf{w}_a^T[\mathbf{h}; 1] \\
\hat{y}_a &= \text{sum}([\mathbf{q}_a\odot\mathbf{q}_a;b_a])
\end{aligned}
\label{eq:fm}
\end{equation}
where $\mathbf{U}_a \in \mathbb{R}^{k \times d}, \mathbf{w}_a \in \mathbb{R}^{d+1}$, $k$ is the number of factor units, and $a$ denotes the $a$-th class. By denoting $\bar{\mathbf{h}} = [\mathbf{h}; 1]$, we can rewrite the decision function of $\hat{y_a}$ in Eq.~(\ref{eq:fm}) as follows:
\begin{equation}
\small
\begin{aligned}
\hat{y}_a
&= \sum_{f=1}^k \left(\sum_{i=1}^d \mathbf{U}_a(f, i)\mathbf{h}(i)\right)^2 + \sum_{i=1}^{d+1} \mathbf{w}_a(i)\bar{\mathbf{h}}(i) \\
&= \sum_{f=1}^k \left(\sum_{i=1}^d \mathbf{U}_a(f, i)\mathbf{h}(i)\right) \left(\sum_{j=1}^d \mathbf{U}_a(f, j)\mathbf{h}(j)\right) + \sum_{i=1}^{d+1} \mathbf{w}_a(i)\bar{\mathbf{h}}(i) \\
&= \sum_{f=1}^k \sum_{i=1}^d \sum_{j=1}^d \mathbf{U}_a(f, i)\mathbf{U}_a(f, j)\mathbf{h}(i)\mathbf{h}(j) + \sum_{i=1}^{d+1} \mathbf{w}_a(i)\bar{\mathbf{h}}(i) \\
&= \sum_{i=1}^d \sum_{j=1}^d \left<\mathbf{U}_a(:, i),\mathbf{U}_a(:, j)\right>\mathbf{h}(i)\mathbf{h}(j) + \sum_{i=1}^d \mathbf{w}_a(i)\mathbf{h}(i) + \mathbf{w}_a(d+1)
\end{aligned}
\end{equation}

One can easily see that this is similar to the two-way Factorization Machines \cite{rendle2012factorization} except that the subscript $j$ ranges from $i+1$ to $d$ in the original form.

\noindent\textbf{Multi-view Machine layer}.
In contrast to modeling up to the second-order feature interactions between all input units as in the Factorization Machine layer, we could further explore all feature interactions up to the $m$th-order between inputs from $m$ views as shown in Figure~\ref{fig:fig_mvm}.
\begin{equation}
\begin{aligned}
\mathbf{q}_a^{(p)} &= \mathbf{U}_a^{(p)}[\mathbf{h}^{(p)}; 1] \\
\hat{y}_a &= \text{sum}([\mathbf{q}_a^{(1)}\odot\cdots\odot\mathbf{q}_a^{(m)}])
\end{aligned}
\label{eq:mvm}
\end{equation}
where $\mathbf{U}_a^{(p)} \in \mathbb{R}^{k \times (d_h+1)}$ is the factor matrix of the $p$-th view for the $a$-th class. By denoting $\bar{\mathbf{h}}^{(p)} = [\mathbf{h}^{(p)}; 1], p = 1, \cdots, m$, we can verify that Eq.~(\ref{eq:mvm}) is equivalent to Multi-view Machines \cite{cao2016multi}.
\begin{equation}
\small
\begin{aligned}
\hat{y}_a
&= \sum_{f=1}^k \prod_{p=1}^m \left(\sum_{i_p=1}^{d_h+1} \mathbf{U}_a^{(p)}(f, i_p)\bar{\mathbf{h}}^{(p)}(i_p)\right) \\
&= \sum_{f=1}^k \sum_{i_1=1}^{d_h+1} \cdots \sum_{i_m=1}^{d_h+1} \left(\prod_{p=1}^m \mathbf{U}_a^{(p)}(f, i_p)\bar{\mathbf{h}}^{(p)}(i_p)\right) \\
&= \sum_{i_1=1}^{d_h+1} \cdots \sum_{i_m=1}^{d_h+1} \left(\sum_{f=1}^k \prod_{p=1}^m \mathbf{U}_a^{(p)}(f, i_p)\right) \left(\prod_{p=1}^m \bar{\mathbf{h}}^{(p)}(i_p)\right)
\end{aligned}
\end{equation}

As shown in Figure~\ref{fig:architecture}, the full-order feature interactions across multiple views are modeled in a tensor, and they are factorized in a collective manner.

Note that a dropout layer \cite{hinton2012improving} is applied before feeding the output from GRU to the fusion layer which is a regularization method designed to prevent co-adaptation of feature detectors in deep neural networks. The dropout method randomly sets each unit as zero with a certain probability. The dropout units contribute to neither the feed-forward process nor the back-propagation process.

Following the computational graph, it is straightforward to compute gradients for model parameters in both the Factorization Machine layer and the Multi-view Machine layer, as we do for the conventional fully connected layer. Therefore, the error messages generated from the loss function on the final mood score can be back-propagated through these fusion layers all the way to the very beginning, {\em i.e.}, $\mathbf{W}_r$, $\mathbf{U}_r$, $\mathbf{W}_z$, $\mathbf{U}_z$, $\mathbf{W}$, $\mathbf{U}$ in GRU for each input view. In this manner, we can say that DeepMood is an end-to-end learning framework for mood detection.

\section{Experiments}
\label{sec:exp}

We investigate a session-level prediction problem. That is to say, we use features of alphanumeric characters, special characters and accelerometer values in a session to predict the mood score of the associated participant.

\subsection{Experimental Setup}

The implementation is completed using Keras \cite{chollet2015keras} with Tensorflow \cite{tensorflow2015-whitepaper} as the backend. The code has been made available at the author's homepage\footnote{\url{https://www.cs.uic.edu/~bcao1/code/DeepMood.py}}. Specifically, a bidirectional GRU is applied on each view of the metadata. RMSProp \cite{tieleman2012lecture} is used as the optimizer. We truncate sessions that contain more than 100 keypresses, and we remove sessions if any of their views contain less than 10 keypresses. It leaves us with 14,613 total samples which are then split by time for training and validation. Each user contributes first 80\% of her sessions for training and the rest for validation. We empirically set other parameters, including the number of epochs, batch size, learning rate and dropout fraction. The number of recurrent units and factor units are selected on the validation set. Detailed configurations of the hyper-parameters are summarized in Table~\ref{tab:para}.

Experiments on the depression score HDRS are conducted as a binary classification task where $c=2$. We consider sessions with the HDRS score between 0 and 7 (inclusive) as negative samples (normal) and those with HDRS greater than or equal to 8 as positive samples (from mild depression to severe depression). On the other hand, the mania score YMRS is more complicated without a widely adopted threshold. Therefore, YMRS is directly used as the label for a regression task where $c=1$. Accuracy and F-score are used to evaluate the classification task, and root-mean-square error (RMSE) is used for the regression task.

\begin{table}[t]
\centering
\caption{Parameter configuration.}
\label{tab:para}
\begin{tabular}{||l|c||}
\hline
Parameter & Value \\
\hline\hline
\# recurrent units ($d_h$)  & 4, 8, 16 \\
\# factor units ($k$)       & 4, 8, 16 \\
\# epochs                   & 500 \\
batch size                  & 256 \\
learning rate               & 0.001 \\
dropout fraction            & 0.1 \\
maximum sequence length     & 100 \\
minimum sequence length     & 10 \\
\hline
\end{tabular}
\end{table}

\subsection{Compared Methods}

The compared methods are summarized as follows:
\begin{itemize}[leftmargin=*,noitemsep,topsep=0pt]
\item\textbf{\dmvm}: The proposed DeepMood architecture with a Multi-view Machine layer for data fusion.
\item\textbf{\dfm}: The proposed DeepMood architecture with a Factorization Machine layer for data fusion.
\item\textbf{\dnn}: The proposed DeepMood architecture with a conventional fully connected layer for data fusion.
\item\textbf{\xgb}: The implementation of a tree boosting system from \texttt{XGBoost}\footnote{\url{https://github.com/dmlc/xgboost}} \cite{chen2016xgboost} is used. We concatenate the sequence data with the maximum length 100 (padding 0 for short ones) of each feature as the input.
\item\textbf{\svm} and \textbf{\lr}: These are two linear models. With the same input setting as {\xgb}, the implementations of Linear Support Vector Classification/Regression and Logistic/Ridge Regression from \texttt{scikit-learn}\footnote{\url{http://scikit-learn.org}} are used for Classification/Regression tasks.
\end{itemize}

In general, {\dmvm}, {\dfm} and {\dnn} can be categorized as late fusion approaches, while {\xgb}, {\svm} and {\lr} are early fusion strategies for the sequence prediction problem on multi-view time series. Note that the number of model parameters for fusing multi-view data in {\dmvm} and {\dfm} is $ck(d+m)$ and $ckd + c(d+1)$, respectively, thereby leading to approximately the same model complexity $O(ckd)$ due to $m \ll d$.
For {\dnn}, the number of model parameters for fusion is $ck' + k'(d+1)$. For a fair comparison, we need to control the model complexity of the compared methods at the same level. Therefore, in all experiments, we always set $k' = ck$.

\subsection{Prediction Performance}

Experimental results are shown in Table~\ref{tab:mainresult}. We can see that the late fusion based DeepMood methods are the best on the prediction for the dichotomized HDRS scores, especially {\dmvm} and {\dfm} with 90.31\% and 90.21\%, respectively. It demonstrates the feasibility of using passive typing dynamics from mobile phone metadata to predict the disturbance and severity of mood states. In addition, it is found that {\svm} and {\lr} are not a good fit to this task, or sequence prediction in general. {\xgb} performs reasonably well as an ensemble method, but {\dmvm} still outperforms it by a significant margin 5.56\%, 5.93\% and 10.02\% in terms of accuracy, F-score and RMSE, respectively. Among the DeepMood variations, the improvement of {\dmvm} and {\dfm} over {\dnn} reveals the potential of replacing a conventional fully connected layer with a Multi-view Machine layer or Factorization Machine layer for data fusion in a deep framework. This is because {\dmvm} and {\dfm} can explicitly capture higher order interactions among features,  while {\dnn} does not capture any feature interaction.

\begin{table}[t]
\caption{Prediction performance of compared methods.}
\small
\label{tab:mainresult}
\centering
\begin{tabular}{||l|c|c|c||}
\hline
Task & \multicolumn{2}{c|}{Classification} & Regression \\
\hline
Metric & Accuracy & F-score & RMSE \\
\hline\hline
{\dmvm}	& 0.9031 & 0.9070 & 3.5664 \\
{\dfm}	& 0.9021 & 0.9029 & 3.6767 \\
{\dnn}	& 0.8868 & 0.8929 & 3.7874 \\
{\xgb}	& 0.8555 & 0.8562 & 3.9634 \\
{\svm}	& 0.7323 & 0.7237 & 4.1257 \\
{\lr}	& 0.7293 & 0.7172 & 4.1822 \\
\hline
\end{tabular}
\end{table}

\begin{figure}[t]
\centering
\begin{minipage}[l]{0.8\columnwidth}
\centering
\includegraphics[width=1\textwidth]{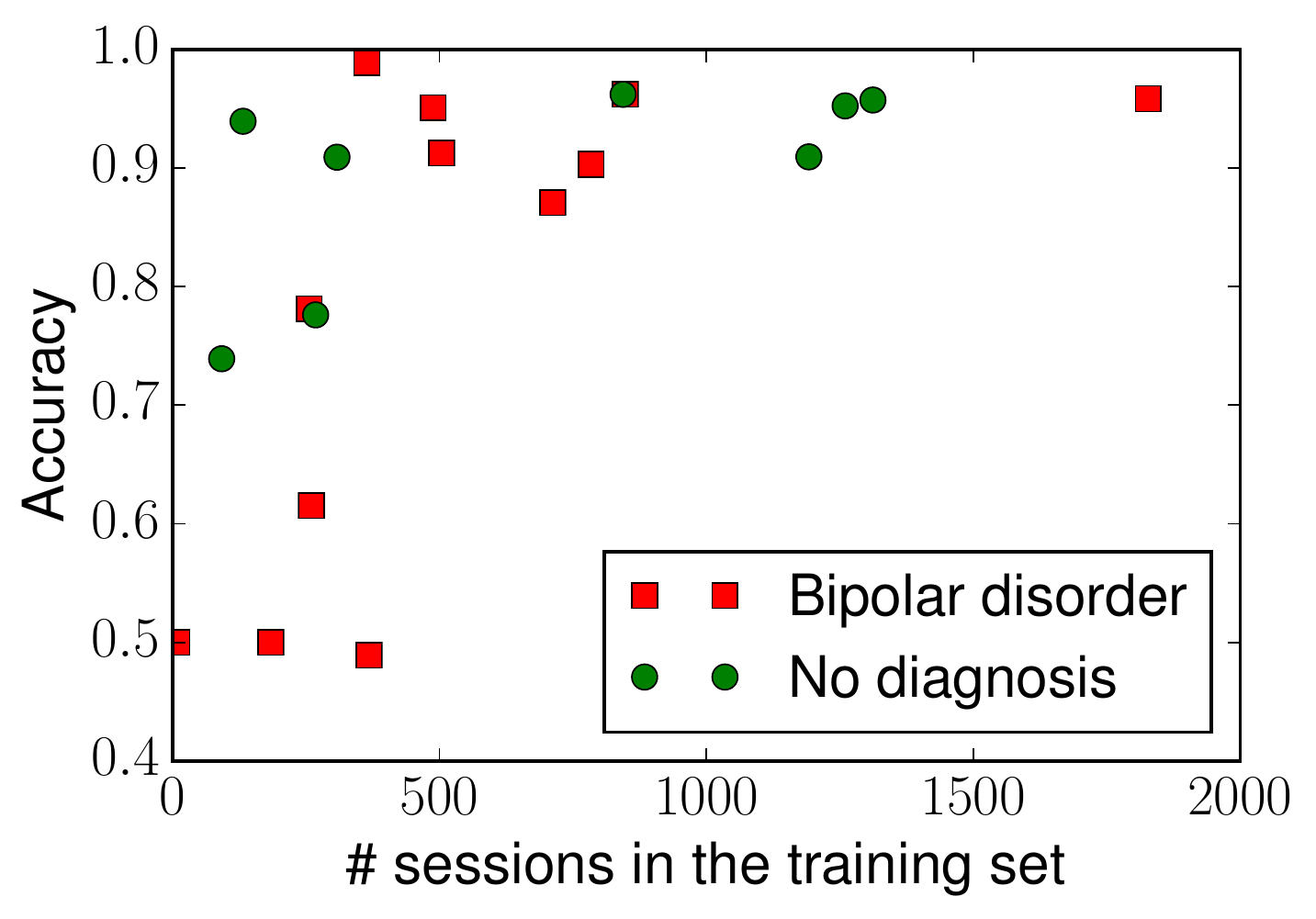}
\end{minipage}
\caption{Prediction performance of {\dmvm} on each of the 20 participants.}
\label{fig:acc_per_subj}
\end{figure}

In practice, it is important to understand how the model works for each individual when monitoring her mood states. Therefore, we investigate the prediction performance of {\dmvm} on each of the 20 participants in our dataset. Results are shown in Figure~\ref{fig:acc_per_subj} where each dot represents a participant with the number of her contributed sessions in the training set and the corresponding prediction accuracy. We can see that the proposed model can steadily produce accurate predictions ($\ge$87\%) of a participant's mood states when she provides more than 400 valid typing sessions in the training phase. Note that the prediction we make in this work is per session which is typically less than one minute. We can expect more accurate results on the daily level by ensembling sessions occurring during a day.

\subsection{Convergence Efficiency}

\begin{figure}[t]
\centering
\begin{minipage}[l]{0.8\columnwidth}
\centering
\includegraphics[width=1\textwidth]{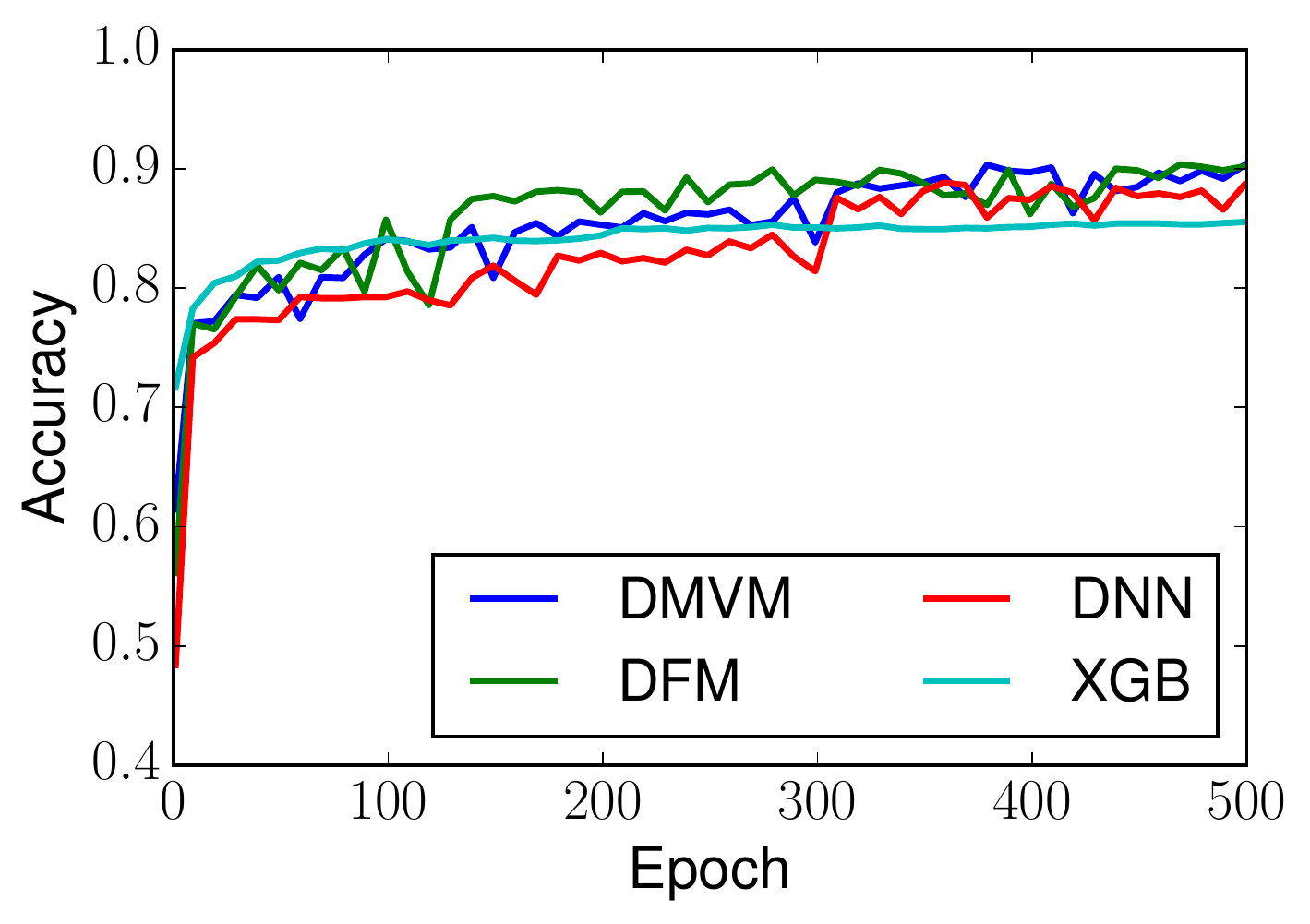}
\end{minipage}
\caption{Learning procedure.}
\label{fig:epoch}
\end{figure}

In this section, we show more details about the learning procedure of the proposed DeepMood architecture with different fusion layers and that of {\xgb}. Figure~\ref{fig:epoch} illustrates how the accuracy on the validation set changes over epochs. We observe that different fusion layers have different convergence performance in the first 300 epochs, and afterwards they steadily outperform {\xgb}. Among the DeepMood methods, it is found that {\dmvm} and {\dfm} converge more efficiently than {\dnn} in the first 300 epochs, and they reach a better local minima of the loss function at the end. This again shows the importance of the fusion layer in a deep framework. It is also interesting to see the convergence process of {\xgb} considering its popularity and success on many tasks in practice.
We found that the generalizability of {\xgb} on the sequence prediction task is limited, although its training error could perfectly converge to 0 at an early stage.

\subsection{Importance of Different Views}

To better understand the role that different views play in the buildup of mood detection by DeepMood, we examine separate models trained with or without each view. Since {\dmvm} is designed for heterogeneous data fusion, {\em i.e.}, data with at least two views, we train {\dmvm} on every pairwise views. Moreover, we train {\dfm} on every single view. Experimental results are shown in Table~\ref{tab:subsetresult}. First, we observe that {\nonch} are poor predictors of mood states. {\ch} and {\accel} have significantly better predictive performance. {\ch} are the best individual predictors of mood states. It validates a high correlation between the mood disturbance and typing patterns including duration of a keypress, time interval since the last keypress, as well as accelerometer values.

\begin{table}[t]
\caption{Prediction performance using different views of the metadata.}
\small
\label{tab:subsetresult}
\centering
\begin{tabular}{||l|c|c|c||}
\hline
Task & \multicolumn{2}{c|}{Classification} & Regression \\
\hline
Metric & Accuracy & F-score & RMSE \\
\hline\hline
{\dmvm} w/o {\ch}	& 0.8125 & 0.8164 & 3.9833 \\
{\dmvm} w/o {\nonch}	& 0.9008 & 0.9034 & 3.8166 \\
{\dmvm} w/o {\accel}	& 0.8318 & 0.8253 & 3.9499 \\
\hline
{\dmvm} w/ all	& 0.9031 & 0.9070 & 3.5664 \\
\hline\hline
{\dfm} w/ {\ch}	& 0.8322 & 0.8224 & 3.9515 \\
{\dfm} w/ {\nonch}	& 0.6260 & 0.5676 & 4.1040 \\
{\dfm} w/ {\accel}	& 0.8015 & 0.8089 & 3.9722 \\
\hline
{\dfm} w/ all	& 0.9021 & 0.9011 & 3.6767 \\
\hline
\end{tabular}
\end{table}



\section{Related Work}

This work is studied in the context of supervised sequence prediction. Xing et al.~provide a brief survey on the sequence prediction problem where sequence data are categorized into five subtypes: simple symbolic sequences, complex symbolic sequences, simple time series, multivariate time series, and complex event sequences \cite{xing2010brief}. Sequence classification methods are grouped into three subtypes: feature based methods, sequence distance based methods, and model based methods. Feature based methods first transform a sequence into a feature vector and then apply conventional classification models \cite{lesh1999mining,aggarwal2002effective,leslie2004fast,ji2007mining,ye2009time}. Distance based methods include K nearest neighbor classifier \cite{keogh2000scaling,keogh2003need,ratanamahatana2004making,wei2006semi,xi2006fast,ding2008querying} and SVM with local alignment kernel \cite{lodhi2002text,she2003frequent,sonnenburg2005large} by measuring the similarity between a pair of sequences. Model based methods assume that sequences in a class are generated by an underlying probability distribution, including Naive Bayes \cite{cheng2005protein}, Markov Model \cite{yakhnenko2005discriminatively} and Hidden Markov Model \cite{srivastava2007hmm}.

However, most of the works focus on simple symbolic sequences and simple time series, with a few on complex symbolic sequences and multivariate time series. The problem of classifying complex event sequence data (a combination of multiple numerical measurements and categorical fields) still needs further investigation which motivates this work. Furthermore, most of the methods are devoted to shallow models with feature engineering. Inspired by the great success of deep RNNs in the applications of other sequence tasks, including speech recognition \cite{graves2013speech} and natural language processing \cite{mikolov2010recurrent,bahdanau2014neural}, in this work, we propose a deep architecture to model complex event sequences of mobile phone typing dynamics.

On multi-view learning, Cao et al.~propose to fuse multi-view data through the operation of tensor product and assume that the effects of feature interactions across views have a low rank \cite{cao2014tensor,cao2016multi}. Lu et al.~extend it to multi-task learning \cite{lu2017multilinear}. Zhang et al.~use Factorization Machines to initialize the bias terms and embedding vectors for multi-field categorical data at the bottom layer of a deep architecture \cite{zhang2016deep}. There are also some work incorporating multiple views into the process of subgraph mining \cite{cao2015mining} and deep learning \cite{zhang2016identifying} to help identify meaningful patterns from data. 

\section{Conclusion}

It appears that mobile phone metadata could be used to predict the presence of mood disorders. The proposed DeepMood architecture is able to achieve 90.31\% prediction accuracy, where late fusion is indeed more effective than early fusion and more sophisticated fusion layer also helps. The ability to passively collect data that can be used to infer the presence and severity of mood disturbances may enable providers to provide interventions to more patients earlier in their mood episodes. Models such as the one presented here may also lead to deeper understanding of the effects of mood disturbances in the daily activities of people with mood disorders.

\section{Acknowledgements}
This work is supported in part by NSF through grants IIS-1526499, and CNS-1626432, and NSFC 61672313.


\bibliographystyle{ACM-Reference-Format}
\bibliography{references}

\end{document}